\documentclass[lettersize,journal]{IEEEtran}
\usepackage{cite}
\usepackage{amsmath,amssymb,amsfonts}
\usepackage{graphicx}
\usepackage{textcomp}
\usepackage[table]{xcolor}
\usepackage{amsmath,amsfonts}
\usepackage{algpseudocode}
\usepackage{algorithm}
\usepackage{array}
\usepackage[caption=false,font=normalsize,labelfont=sf,textfont=sf]{subfig}
\usepackage{textcomp}
\usepackage{stfloats}
\usepackage{url}
\usepackage{tcolorbox}
\usepackage{xurl}
\usepackage{verbatim}
\usepackage{graphicx}
\usepackage{booktabs}
\usepackage{xspace}
\usepackage{listings}
\usepackage{hyperref}
\usepackage{footmisc}
\usepackage{caption}
\usepackage{booktabs} 
\definecolor{dkgreen}{rgb}{0,0.6,0}
\definecolor{gray}{rgb}{0.5,0.5,0.5}
\definecolor{mauve}{rgb}{0.58,0,0.82}
\definecolor{lightgray}{rgb}{0.95,0.95,0.95} 
\lstdefinelanguage{Solidity}{
  keywords=[1]{anonymous, assembly, assert, balance, break, call, callcode, case, catch, class, constant, continue, constructor, contract, debugger, default, delegatecall, delete, do, else, emit, event, experimental, export, external, false, finally, for, function, gas, if, implements, import, in, indexed, instanceof, interface, internal, is, length, library, log0, log1, log2, log3, log4, memory, modifier, new, payable, pragma, private, protected, public, pure, push, require, return, returns, revert, selfdestruct, send, solidity, storage, struct, suicide, super, switch, then, this, throw, transfer, true, try, typeof, using, value, view, while, with, addmod, ecrecover, keccak256, mulmod, ripemd160, sha256, sha3}, 
  keywordstyle=[1]\color{blue}\bfseries,
  keywords=[2]{address, bool, byte, bytes, bytes1, bytes2, bytes3, bytes4, bytes5, bytes6, bytes7, bytes8, bytes9, bytes10, bytes11, bytes12, bytes13, bytes14, bytes15, bytes16, bytes17, bytes18, bytes19, bytes20, bytes21, bytes22, bytes23, bytes24, bytes25, bytes26, bytes27, bytes28, bytes29, bytes30, bytes31, bytes32, enum, int, int8, int16, int24, int32, int40, int48, int56, int64, int72, int80, int88, int96, int104, int112, int120, int128, int136, int144, int152, int160, int168, int176, int184, int192, int200, int208, int216, int224, int232, int240, int248, int256, mapping, string, uint, uint8, uint16, uint24, uint32, uint40, uint48, uint56, uint64, uint72, uint80, uint88, uint96, uint104, uint112, uint120, uint128, uint136, uint144, uint152, uint160, uint168, uint176, uint184, uint192, uint200, uint208, uint216, uint224, uint232, uint240, uint248, uint256, var, void, ether, finney, szabo, wei, days, hours, minutes, seconds, weeks, years},  
  keywordstyle=[2]\color{teal}\bfseries,
  keywords=[3]{block, blockhash, coinbase, difficulty, gaslimit, number, timestamp, msg, data, gas, sender, sig, value, now, tx, gasprice, origin},  
  keywordstyle=[3]\color{violet}\bfseries,
  identifierstyle=\color{black},
  sensitive=false,
  comment=[l]{//},
  morecomment=[s]{/*}{*/},
  commentstyle=\color{gray}\ttfamily,
  stringstyle=\color{red}\ttfamily,
  morestring=[b]',
  morestring=[b]"
}

\usepackage{color}
\usepackage[T1]{fontenc}
\usepackage{fancyhdr}

\newcommand{\boxmargin}{1mm}
\newtcolorbox{myboxc}{
    colback=gray!15!white,
    arc = 0pt, outer arc = 0pt,
    boxsep=0pt, left = 3pt, right = 0pt, top = 0pt, bottom = 0pt, 
    leftrule=3pt, bottomrule=0pt,toprule=0pt, rightrule=0pt,
    left = \boxmargin, right = \boxmargin, top = \boxmargin, bottom = \boxmargin
}
\algrenewcommand\algorithmicrequire{\textbf{Input:}}
\algrenewcommand\algorithmicensure{\textbf{Output:}}

\newcommand{\ac}{\textit{Access Control}\xspace}
\newcommand{\toolname}{\textsc{Trace}\xspace}
\newcommand{\gptfouro}{\textit{gpt-4o}\xspace}

\def\BibTeX{{\rm B\kern-.05em{\sc i\kern-.025em b}\kern-.08em
    T\kern-.1667em\lower.7ex\hbox{E}\kern-.125emX}}
    
\begin{document}

\title{\textsc{Trace}: Securing Smart Contract Repository Against Access Control Vulnerability}

\author{
    Chong Chen, 
    Lingfeng Bao, 
    David Lo,~\IEEEmembership{Fellow,~IEEE},
    Yanlin Wang, 
    Zhenyu Shan, 
    Ting Chen, 
    Guangqiang Yin, 
    Jianxing Yu, 
    Zibin Zheng,~\IEEEmembership{Fellow,~IEEE},
    Jiachi Chen
    \thanks{Chong Chen, Yanlin Wang, Jianxing Yu, Zibin Zheng, Jiachi Chen are with the School of Software Engineering, Sun Yat-sen University, Zhuhai 519082, China (e-mail:  \href{mailto:chench578@mail2.sysu.edu.cn}{chench578\allowbreak@mail2\allowbreak.sysu\allowbreak.edu\allowbreak.cn}; \href{mailto:wangylin36@mail.sysu.edu.cn}{wangylin36\allowbreak@mail\allowbreak.sysu\allowbreak.edu\allowbreak.cn}; \href{mailto:yujx26@mail.sysu.edu.cn}{yujx26\allowbreak@mail\allowbreak.sysu\allowbreak.edu\allowbreak.cn}; \href{mailto:zhzibin@mail.sysu.edu.cn}{zhzibin\allowbreak@mail\allowbreak.sysu\allowbreak.edu\allowbreak.cn};\href{mailto:chenjch86@mail.sysu.edu.cn}{chenjch86\allowbreak@mail\allowbreak.sysu\allowbreak.edu\allowbreak.cn};)
    }
    \thanks{Lingfeng Bao is with the College of Computer Science and Technology, Zhejiang University, Hangzhou 310013, China (e-mail:  \href{mailto:lingfengbao@zju.edu.cn}{lingfengbao\allowbreak@zju\allowbreak.edu\allowbreak.cn})}
    \thanks{David Lo is with the School of Computing and Information Systems, Singapore Management University, Singapore (e-mail:  \href{mailto:davidlo@smu.edu.sg}{davidlo\allowbreak@smu\allowbreak.edu\allowbreak.sg})}
    \thanks{Zhenyu Shan is with the Intelligent Transportation and Information Security Laboratory, Hangzhou Normal University, Hangzhou 311121, China (e-mail: \href{mailto:20100119@hznu.edu.cn}{20100119\allowbreak@hznu\allowbreak.edu\allowbreak.cn})}
    \thanks{Ting Chen, Guangqiang Yin are with the School of Computer Science and Engineering (School of Cyber Security), University of Electronic Science and Technology of China, Chengdu 611731, China, and also with Kashi Institute of Electronics and Information Industry, Kashi, 844000, China (e-mail: \href{mailto:brokendragon@uestc.edu.cn}{brokendragon\allowbreak@uestc\allowbreak.edu\allowbreak.cn}; \href{mailto:yingq@uestc.edu.cn}{yingq\allowbreak@uestc\allowbreak.edu\allowbreak.cn})
}
}

\maketitle

\begin{abstract}
Smart contract vulnerabilities have led to billions of dollars in economic losses. Among these, improper \ac, which allows unauthorized users to execute restricted functions, is particularly prevalent and has caused significant financial damage. 
Smart contract repositories contain source code, documentation, configuration files, and other artifacts necessary for building and deploying smart contracts. GitHub hosts numerous open-source repositories of this kind, which serve as intermediate artifacts in development and require compilation and packaging to produce deployable contracts. Third-party developers often reference, reuse, or fork code from these repositories during custom development. However, if the referenced code contains vulnerabilities, it can introduce significant security risks. Existing tools for detecting smart contract vulnerabilities are limited in their ability to handle such complex repositories, as they typically require the target contract to be compilable to generate an abstract representation of the program for further analysis.
This paper presents \toolname, a tool designed to secure non-compilable smar\underline{t} contract \underline{r}epositories against \underline{a}ccess \underline{c}ontrol vuln\underline{e}rabilities. \toolname employs LLMs to locate sensitive functions involving critical operations (e.g., \textit{transfer}) within the contract and subsequently completes function snippets into a fully compilable contract. 
\toolname constructs a function call graph from the abstract syntax tree (AST) of the completed contract. It uses the control flow graph (CFG) of each function as node information. The nodes of the sensitive functions are then analyzed to detect \ac vulnerabilities. Experimental results demonstrate that \toolname outperforms state-of-the-art tools on an open-sourced CVE dataset, detecting 14 out of 15 CVEs. In addition, it achieves 89.2\% precision on 5,000 recent on-chain contracts, far exceeding the best existing tool at 76.9\%. On 83 real-world repositories, \toolname achieves 87.0\% precision, significantly surpassing \textit{DeepSeek-R1}’s 14.3\%.
\end{abstract}


\section{Introduction}
Smart contracts are self-executing programs that automatically enforce agreements without intermediaries~\cite{chen2021defectchecker}. These contracts operate on decentralized platforms, offering a mechanism for automating transactions and interactions. As a result, they form the backbone of decentralized applications (DApps)~\cite{cai2018decentralized}, enabling functionality such as financial transactions~\cite{chitta2019decentralized}, supply chain tracking~\cite{muller2019hidals}, and digital identity management~\cite{satybaldy2022decentralized}. By leveraging smart contracts, DApps provide users with trustless, transparent, and immutable services, eliminating reliance on centralized entities.

Despite their advantages, smart contracts face significant security challenges, particularly \ac vulnerability~\cite{ac}. These issues arise when contracts fail to properly restrict or validate permissions for executing sensitive functions. 
For example, the \textit{Parity Multi-sig bugs} resulted in a loss of 153,037 Ether (valued at 30 million USD at the time)~\cite{paritybugs} due to the improper validation of the $msg.sender$’s permissions, which allowed unauthorized users to assume ownership and execute critical operations. 


Existing tools for detecting \ac vulnerabilities in smart contracts primarily rely on static and dynamic analysis. Static analysis examines the source code to identify potential issues~\cite{slither,oyente,smartcheck,securify}, while dynamic analysis observes contract behavior in controlled execution environments~\cite{mossberg2019manticore}. 
However, these tools are primarily designed for analyzing compilable contracts, assuming the availability of a functioning build system. In practice, many open-source smart contract repositories, especially those still under development, archived, or community-maintained, often lack the necessary configuration files, dependency versions, or coherent structures required for successful compilation. This limitation severely restricts the scope of vulnerability detection, leaving large parts of the contract ecosystem unanalyzed.

\textbf{Third-party developers} frequently reuse or fork open-source smart contract repositories from GitHub as a foundation for their own DApps. Yet, these repositories often pose challenges for \textbf{automated security analysis}, as most existing tools rely on \textbf{compiled artifacts} such as control flow graphs (CFGs) or abstract syntax trees (ASTs) to perform accurate detection. Many of the repositories lack standardized build instructions and contain complex, interdependent contracts written in varying versions of \textit{Solidity}~\cite{solidity}. This makes it difficult to compile, resolve dependencies, and analyze the code without extensive manual setup. Although these repositories serve as developmental artifacts—including source code, documentation, and configurations—their raw form is often incomplete or inconsistent, especially in early-stage or inactive projects. Assessing their security properties can be non-trivial, hindering safe reuse by other developers and delaying secure integration into larger systems.

As outlined in Section~\ref{subsec:lifecycle}, the compilation step is often complex and resource-intensive, which may hinder the scalability and availability of security auditing. Large language models (LLMs) have shown promise in tasks like code generation, summarization, and completion. However, a previous study found that directly applying LLMs to vulnerability detection often results in low precision. Additionally, the lengthy code in smart contract repositories can overwhelm LLMs, causing distractions from irrelevant context and reducing performance. Some research methods have attempted to combine LLMs with program analysis for smart contract vulnerability detection. \textit{GPTScan}~\cite{10.1145/3597503.3639117}, a tool for detecting logical vulnerabilities, combines LLMs with program analysis. It uses LLMs to identify potential vulnerability scenarios and then verifies them through program analysis. However, it still requires compiling the entire smart contract project, making it time-consuming. Additionally, \textit{GPTScan} mainly targets logical vulnerabilities rather than \ac issues.

To address these limitations, we propose \toolname, a tool designed to secure non-compilable smar\underline{\textbf{t}} contract \underline{\textbf{r}}epositories against \underline{\textbf{a}}ccess \underline{\textbf{c}}ontrol vuln\underline{\textbf{e}}rabilities. Specifically, \toolname first filters out smart contracts within projects that involve sensitive operations (i.e., selfdestruct, transfer, state variables modification, low-level external contract call). 
To mitigate the impact of irrelevant code, our approach begins by extracting function snippets that contain these sensitive operations. This extraction is performed using an LLM, leveraging its capability for semantic code analysis. Next, \toolname leverages the LLM to complete the sensitive function snippets into complete smart contracts, enabling their compilation to generate intermediate representations for more precise analysis. If the completed contract remains uncompilable, \toolname submits the error details and source code to the LLM for self-reflection, prompting it to revise the code based on the errors. 
\toolname constructs a function call graph (FCG) for the completed contracts, where each function is represented as a node and edges denote function calls. Each node is annotated with sensitive tags, indicating whether the function performs sensitive operations. Additionally, \toolname uses \textit{Slither}~\cite{slither} to generate a function-level control flow graph (CFG) for each node in the FCG. We then defined four types of risky actions that require strict access control and conducted a search for these actions within the FCG. If any such pattern was identified and found to lack proper access control mechanisms, \toolname flagged the presence of a vulnerability.

We evaluated the performance of \toolname using three datasets. The first dataset is derived from related studies~\cite{ghaleb2023achecker,spcon}, which consists of 15 smart contracts with known \ac vulnerability, each associated with a CVE. \toolname successfully identified 14 out of 15 CVEs, outperforming SOTA tool \textit{AChecker}~\cite{ghaleb2023achecker}, which detected 12 out of 15 CVEs. The second dataset includes 5,000 recent smart contracts from the Ethereum~\cite{ethereum}. For this evaluation, we compared \toolname against six baseline tools capable of detecting \ac vulnerability: Slither~\cite{slither}, GPTScan~\cite{10.1145/3597503.3639117}, AChecker~\cite{ghaleb2023achecker}, Securify~\cite{securify}, Mythril~\cite{mythril}, Maian~\cite{maian}, and Manticore~\cite{mossberg2019manticore}. \toolname achieved a precision of 89.2\%, surpassing all the baseline tools, where the highest precision is 55.9\%, with an average of 25.1\%. The third dataset comprises 83 real-world smart contract repositories sampled from the \textit{DAppSCAN} dataset~\cite{zheng2024dappscan}, containing a total of 3,092 smart contracts. \toolname detected 40 \ac vulnerabilities, achieving a precision of 87.0\%. This performance significantly outperforms the best-performing baseline model, \textit{DeepSeek-R1}, which achieved a precision of 14.3\%.

Overall, we have made the following contributions.
\begin{itemize}
\item To the best of our knowledge, this study is the first to detect \ac vulnerabilities in non-compilable smart contract repositories, while the proposed methods remain applicable to independent smart contracts.
\item By integrating LLMs with traditional program analysis techniques, we developed \toolname and evaluated its performance across three datasets from different sources.
\item All source code and data related to this study are publicly available at \texttt{\url{https://github.com/BugmakerCC/Trace}} to support future research.
\end{itemize}

\section{Background}
\label{sec:bkg}

\subsection{Smart Contracts}
\label{subsec:lifecycle}
Smart contracts are autonomous programs that are deployed and executed on blockchain~\cite{chen2021defectchecker}. 
During development and maintenance, \textit{\textbf{smart contract repositories}} are stored and collaboratively developed on platforms like GitHub, which include \textit{\textbf{single smart contract files}}, configuration files, and libraries. Each single smart contract file in the repository implements specific functions that are essential to the overall functionality of the entire repository. Once packaged, compiled, and deployed, these projects become \textit{\textbf{smart contracts deployed on the blockchain}}, where their bytecode is stored under a specific address. \textit{\textbf{Decentralized Application (DApp)}} is a software application that runs on blockchain networks, offering users decentralized and trustless services without relying on centralized authorities~\cite{cai2018decentralized}. Smart contracts deployed on the blockchain serve as the backend of DApps. This lifecycle reflects the evolving roles of smart contracts from development to deployment. 

\begin{figure}[ht]
\centering
\includegraphics[width=\linewidth]{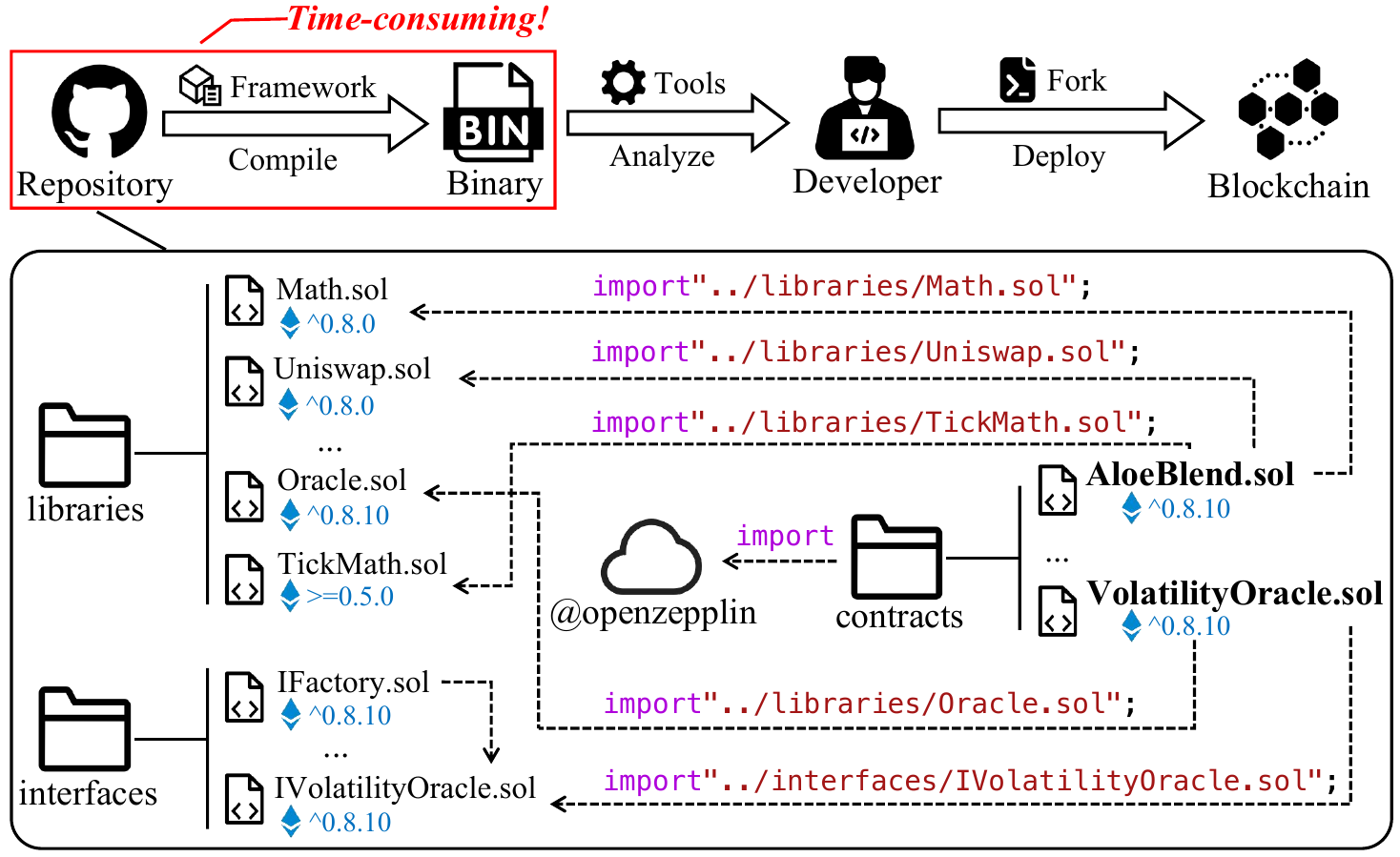}
\caption{The process of compiling, auditing, and forking a smart contract repository.}
\label{fig:repo}
\end{figure}

\subsection{Motivation: Why Analyze Non-Compilable Repositories?}


Real-world smart contract repositories are often fragmented and lack a unified build system. Developers exploring third-party code or researchers auditing historical or incomplete contracts frequently encounter repositories that cannot be compiled due to version conflicts, unresolved dependencies, or deprecated tools. While building frameworks like \textit{Hardhat}~\cite{Hardhat} or \textit{Truffle}~\cite{Truffle} can ease compilation in controlled environments, they are not always present or maintained in community-shared or forked repositories. In our empirical analysis (see~\ref{complexity}), only 6.0\% of repositories from the \textit{DAppSCAN}~\cite{zheng2024dappscan} dataset compiled successfully without manual intervention. This highlights the practical importance of analyzing non-compilable repositories, especially during early-stage audits, third-party integrations, or when revisiting vulnerable legacy code.

Figure~\ref{fig:repo} illustrates the process of compiling, auditing, and forking a smart contract repository as a developer. For the repository in Figure~\ref{fig:repo}, contracts located in the \textit{contracts} folder contain the core logic of the entire project (e.g., \textit{AloeBlend.sol}, \textit{VolatilityOracle.sol}), but they often \textbf{depend on contract files from other folders} (e.g., \textit{AloeBlend.sol} depends on \textit{TickMath.sol} in \textit{libraries} folder). In addition to the interdependencies between files, each file may also be associated with a \textbf{different \textit{Solidity} version}. For instance, \textit{AloeBlend.sol} requires \textit{Solidity} version 0.8.10 or higher, while \textit{TickMath.sol} requires at least version 0.5.0. Therefore, using a single compiler version may lead to syntax errors, necessitating the use of multiple compiler versions and the ability to switch between them flexibly. Additionally, \textbf{external open-source libraries}, such as \textit{OpenZeppelin}~\cite{openzeppelin}, are utilized in some parts of the code. Developers must install and configure these libraries in the correct path. Some \textbf{development frameworks} like \textit{Hardhat}~\cite{Hardhat} or legacy tools like \textit{Truffle}~\cite{Truffle} are also required to package the entire repository and successfully compile it. 
Only after completing this setup can the compiled project be analyzed using security tools, as these tools depend on the intermediate representation of the compiled code to perform security checks. Once a security report is generated, developers can confidently reference and reuse the repository code for custom development tasks, such as forking, before ultimately deploying their finalized contracts on the blockchain. 

\subsection{\ac}
\ac vulnerabilities in smart contracts arise when permission mechanisms fail to restrict unauthorized access to critical operations or resources. Such vulnerabilities can lead to unauthorized fund transfers, manipulation of contract states, or exploitation of administrative privileges. In smart contract repositories, these vulnerabilities can compromise the entire system, causing financial losses and undermining user trust. Ensuring robust access control is crucial for maintaining the reliability of smart contract repositories.

\begin{figure}[ht]
\setlength{\belowcaptionskip}{-0.5cm}
\centering
\includegraphics[width=\linewidth]{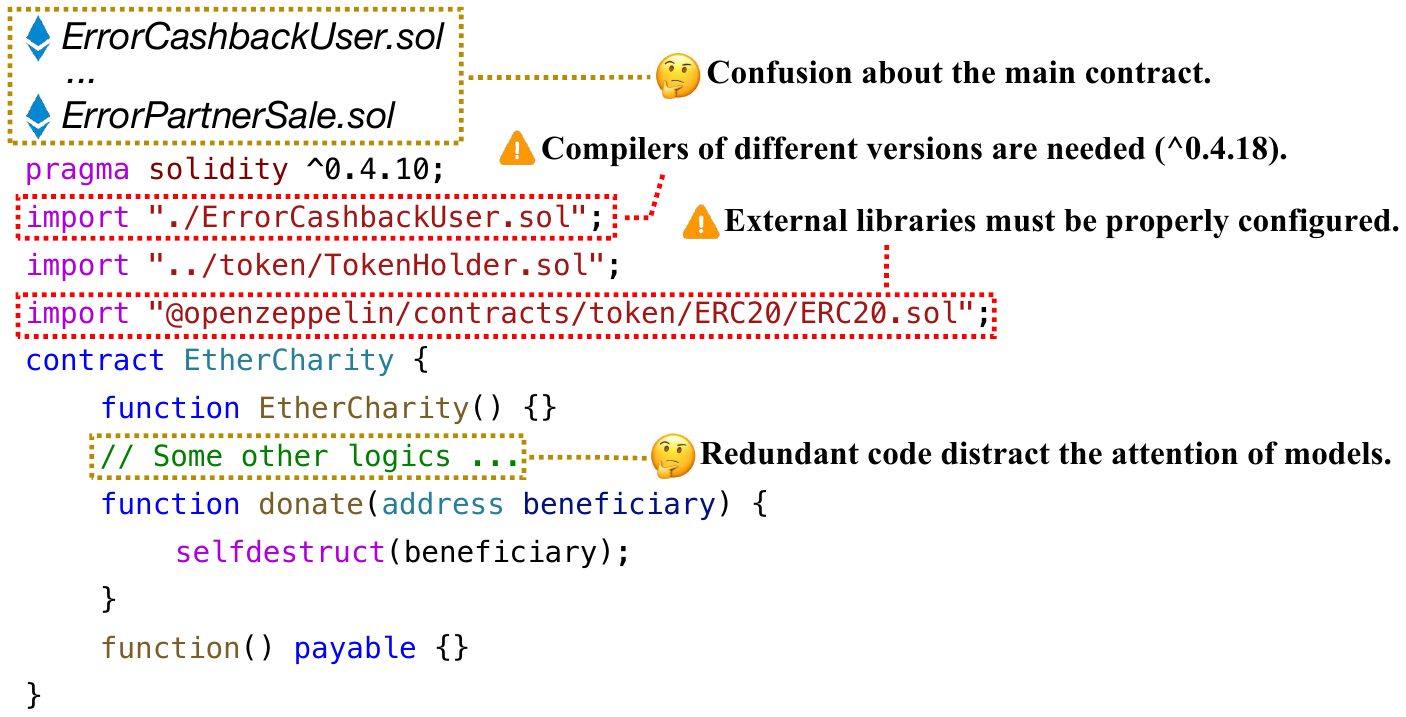}
\caption{An example of \textit{Access Control} vulnerability.}
\label{fig:ac_example}
\end{figure}

Figure~\ref{fig:ac_example} shows an example of \ac vulnerability in a smart contract repository. The $donate$ function in  $EtherCharity$ allows any caller to invoke the $selfdestruct$ operation and transfer the contract’s balance to an arbitrary $beneficiary$ address. This vulnerability arises because the function lacks any form of authorization checks to verify whether the caller is permitted to perform such a critical operation. As a result, anyone can exploit this function by passing their address as the $beneficiary$, draining all funds from the contract. While the vulnerability logic illustrated in Figure~\ref{fig:ac_example} may appear simple, existing tools struggle to detect it effectively. First, a smart contract repository typically consists of multiple contract files, making it challenging for tools to automatically and accurately identify the main contract. Second, contracts often depend on other contract files within the repository, which may require different versions of the \textit{Solidity} compiler. This necessitates developers preparing multiple compiler versions and seamlessly switching between them. Additionally, if the external libraries relied upon by the contract are not properly loaded, compilation errors can arise. Lastly, redundant code within the contract can distract the attention of LLMs or other machine learning methods, resulting in reduced vulnerability detection performance.

\subsection{Static Analysis}

Static analysis is a foundational technique in software security that identifies vulnerabilities by analyzing a program’s code or bytecode without executing it~\cite{staticanalysis}. In smart contracts, static analysis often requires the code to be fully compilable, as it relies on generating bytecode to disassemble into opcodes and simulate execution in the EVM~\cite{evm}. Although some tools generate intermediate representations, such as AST, that do not require contracts to be fully compilable, these representations are often insufficiently precise. They typically remain at the syntactic level and fail to capture the logical information involved during program execution. \textit{Slither}~\cite{slither} is a \textit{Solidity}~\cite{solidity} \& \textit{Vyper}~\cite{vyper} static analysis framework written in Python3. It runs a suite of vulnerability detectors, prints visual information about contract details (e.g., contract summary, evm instructions, state variables). In addition, it provides an API to write custom analyses easily, which means users can customize new vulnerability patterns and integrate them into \textit{Slither} for detection.

\section{The Design of \toolname}
\label{sec:dappguard}

\begin{figure}[h]
\centering
\includegraphics[width=\linewidth]{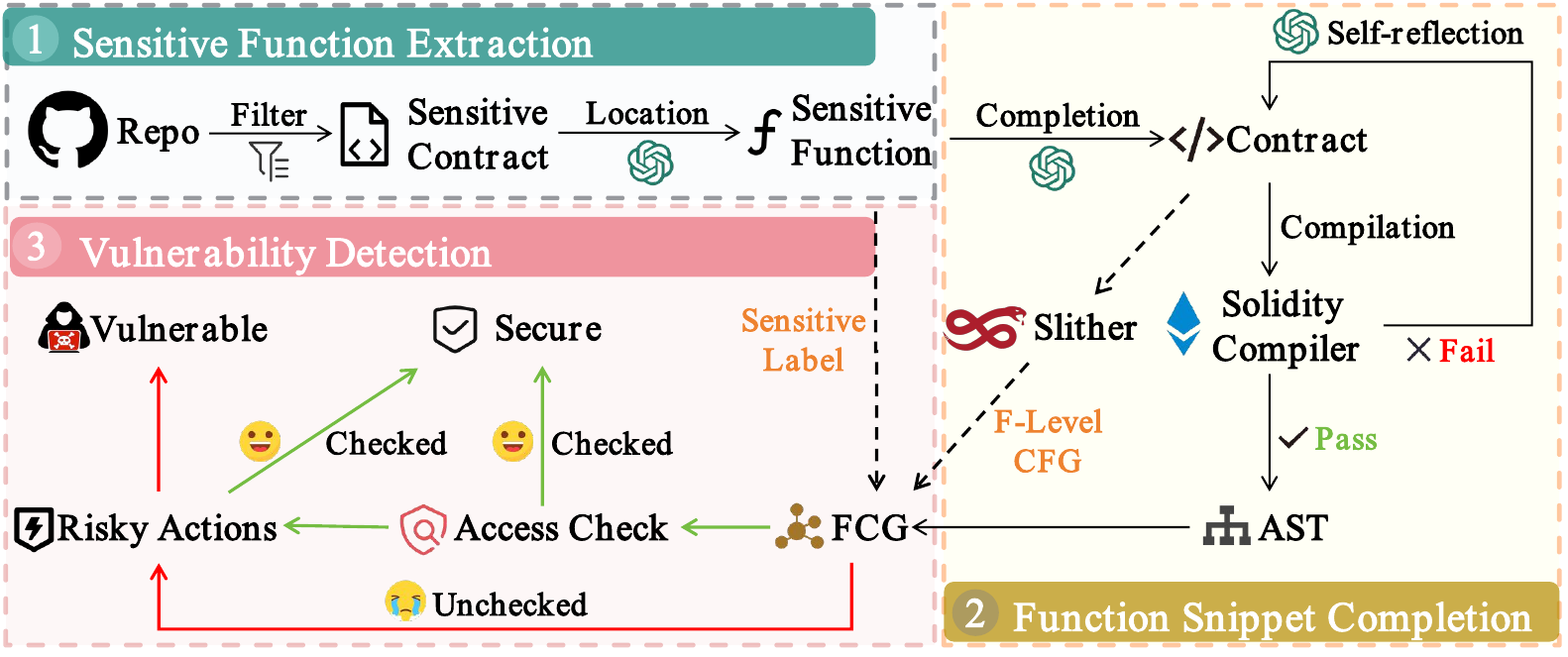}
\caption{The overview of \toolname.}
\label{fig:overview}
\end{figure}

Figure~\ref{fig:overview} illustrates the architecture of \toolname, comprising three interconnected components. The process begins with \textbf{sensitive function extraction}, which identifies functions containing sensitive operations within smart contract repositories, as these are frequent sources of vulnerabilities. The extracted function snippets are then processed through \textbf{function snippet completion}, which transforms them into complete smart contracts while maximizing the likelihood of successful compilation. Finally, the completed contracts undergo \textbf{vulnerability detection}, where static analysis is performed. This involves constructing a function call graph, enriching each function node with intermediate code representations, and evaluating risky actions and their associated access conditions to detect potential \ac vulnerabilities.

\subsection{Sensitive Function Extraction}
\label{subsec:sens_func_extract}
\ac vulnerabilities frequently arise in functions that perform sensitive operations. In this study, we identify four types of such operations, i.e., \textbf{Selfdestruct}, \textbf{Transfer},  \textbf{State Variable Modification}, and \textbf{External Contract Call}. Improper access control of any of these operations can result in serious vulnerabilities. Table~\ref{tab:sens_ops} provides detailed definitions of these operations along with examples of potential vulnerability scenarios. We define the functions that contain these sensitive operations as sensitive functions.


\begin{table*}[!t]
\caption{Definition of sensitive operations and potential vulnerability scenarios.}
\label{tab:sens_ops}
\centering
\renewcommand{\arraystretch}{1} 
\setlength{\tabcolsep}{4pt}       
\begin{tabular}{p{3.4cm}p{5.8cm}p{6.8cm}}
\toprule
\textbf{Sensitive Operations} & \textbf{Definitions} & \textbf{Scenarios} \\
\midrule
Selfdestruct & 
Remove a smart contract from the blockchain permanently, transferring its balance to a specified address. & 
The attacker calls the \texttt{Selfdestruct} function to transfer all balances from the contract to malicious addresses. \\
\midrule
Transfer & 
Send cryptocurrency from a smart contract to a specified address. & 
The attacker transfers the balance from the contract or Externally Owned Account (EOA) to malicious addresses. \\
\midrule
External Contract Call & 
Invoke a function in another contract outside the current contract. & 
The attacker creates an external contract with malicious logic and invokes it in the function. \\
\midrule
State Variable Modification & 
Alter a contract's persistent storage variables, which can impact the contract's behavior and state. & 
The attacker modifies critical state variables in the contract, such as \texttt{balance}, disrupting its ability to operate according to the intended logic. \\
\bottomrule
\end{tabular}
\end{table*}

To focus our analysis on relevant contracts, we prune the search space by excluding files based on their directory paths. Following standard project structures, we omit files from folders like \textit{/interface}, \textit{/library}, \textit{/util}, \textit{/mock}, and \textit{/test}. This step is motivated by the fact that these directories typically house code that is non-critical to production security, such as abstract interfaces, trusted external libraries, and testing apparatus, and are therefore excluded from our analysis.

\begin{figure}[h]
\centering
\includegraphics[width=\linewidth]{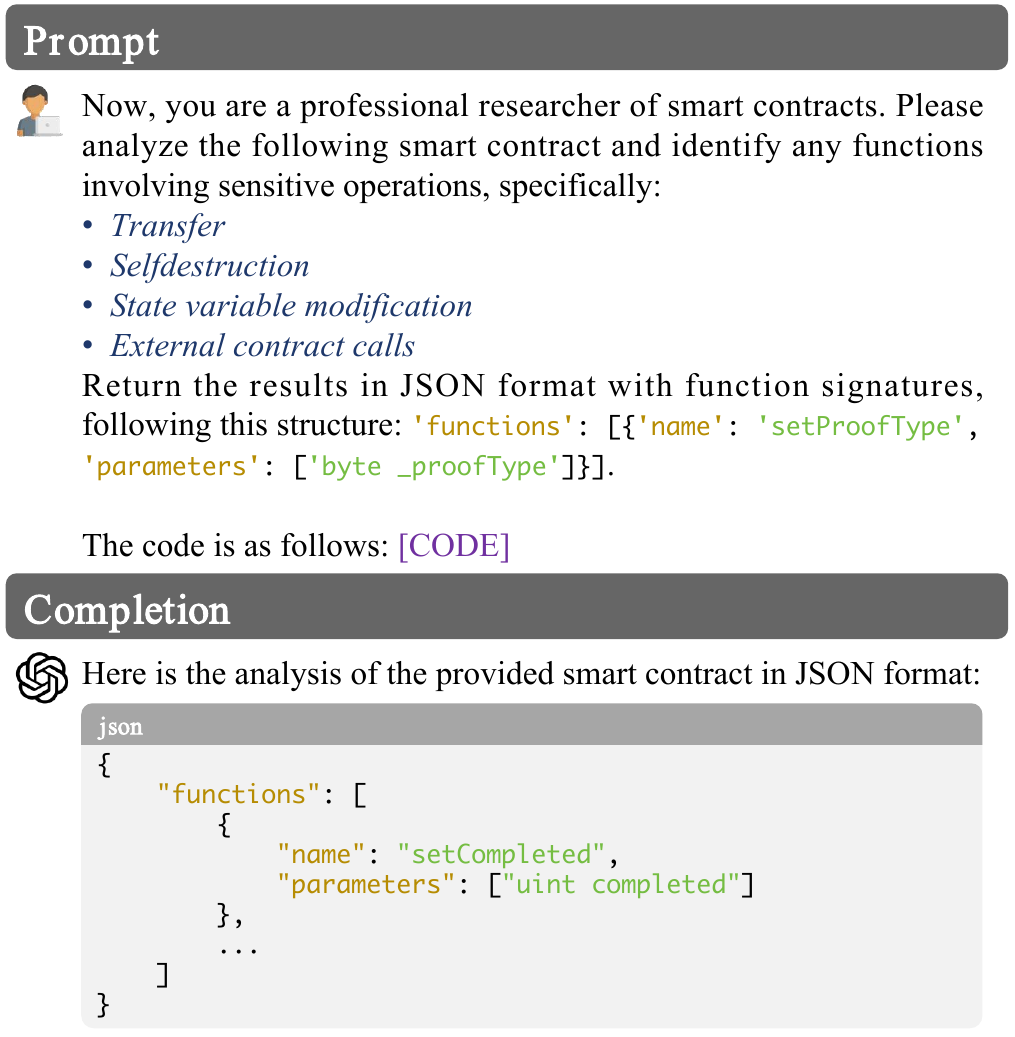}
\caption{Prompt template of sensitive function location.}
\label{fig:sens_func_loc}
\end{figure}

While static analysis is capable of identifying predefined sensitive operations, its efficacy is limited by two key factors. \textit{First}, they depend on a fully compilable environment, which poses a significant obstacle given the complexity resource-intensive of compiling large smart contract repositories (as established in Section~\ref{subsec:lifecycle}). \textit{Second}, static analysis relies on predefined syntactic patterns, which are often difficult to define comprehensively. Accurately specifying all possible variants of sensitive logic is particularly challenging when such functionality is abstracted into external library calls or complex user-defined functions, potentially leading to inaccuracies. In contrast, LLMs overcome both limitations. They operate directly on source code, eliminating the need for compilation, and leverage a deeper semantic understanding of the code's logic rather than pre-defined pattern matching. This allows LLMs to locate sensitive functions with greater accuracy and broader applicability, even in non-compilable contracts. Therefore, we utilize LLMs, specifically the \gptfouro model's API, to perform sensitive function localization. 
Building upon prior work~\cite{chen2024identifying}, we crafted a prompt design as shown in Figure~\ref{fig:sens_func_loc}. 
In the prompt, we specify that the task is to identify sensitive functions within a smart contract and provide a clear definition of what constitutes a sensitive function. Finally, we instruct the model to return the signature of the identified sensitive function. ``[CODE]'' in the Figure~\ref{fig:sens_func_loc} represents the smart contract source code. After obtaining the function signature, we extract the corresponding function from the contract's source code.

\subsection{Function Snippet Completion}

To enable program analysis, we use LLMs to expand extracted sensitive function snippets into complete, compilable contracts. While acknowledging potential complexities from inter-procedural dependencies within the original repository, providing full (and often fragmented or non-compilable) context to the LLM can be counterproductive due to token limits and the risk of model confusion with noisy inputs. To mitigate these issues, we provide the LLM with only the function's raw source code. The LLM then leverages its semantic understanding to infer a minimal and plausible execution context, generating a self-contained, compilable contract. Although this reconstructed contract serves as a localized approximation of the original environment, it critically enables targeted static analysis of the sensitive function's intrinsic logic, which is often intractable in its initially non-compilable state.

\begin{figure}[h]
\centering
\includegraphics[width=\linewidth]{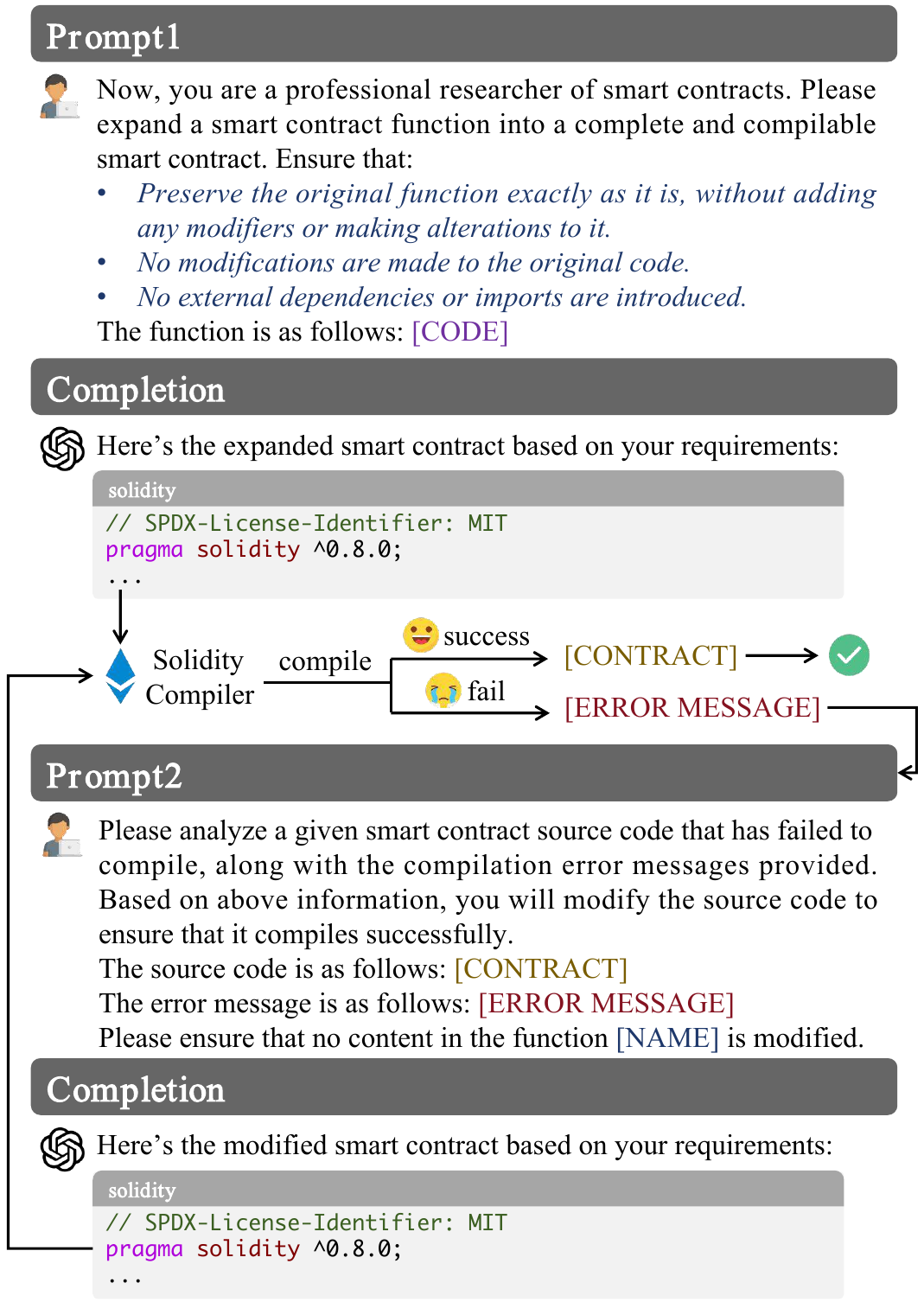}
\caption{Prompt template of function snippet completion.}
\label{fig:sens_func_comp}
\end{figure}

We employ the \gptfouro model, guided by the $Prompt1$ template (Figure~\ref{fig:sens_func_comp}~\cite{chen2024identifying}), to make the sensitive function snippet ``[CODE]'' compilable. To ensure the output accurately reflects the original code's security properties, we impose several key constraints in the prompt. Specifically, the LLM is instructed to refrain from altering the snippet or injecting new logic, as such changes could mask existing vulnerabilities or introduce new ones. This approach produces a self-contained contract that preserves the security context of the original function.

Some code snippets compile successfully on the first attempt, while others encounter compilation errors. To address this, we designed a self-reflection mechanism that enables the LLM to modify the completed code until it compiles successfully iteratively. The prompt template for this process, referred to as $Prompt2$, is shown in Figure~\ref{fig:sens_func_comp}. Specifically, we provide the LLM with the completed contract and the associated compilation error messages, guiding it in fixing the code while ensuring the sensitive function snippets in the source code remain unchanged. In $Prompt2$, ``[CONTRACT]'' represents the completed smart contract, ``[ERROR MESSAGE]'' denotes the compilation error, and ``[NAME]'' indicates the sensitive function's name. After receiving the revised smart contract, we attempt compilation again. If the compilation still fails, the self-reflection process continues as described above.

\subsection{Vulnerability Detection}
The vulnerability detection process is performed individually for each sensitive function identified in the previous steps. That is, each sensitive function snippet is completed into a compilable smart contract, and the following analysis is conducted separately on each of these completed contracts.

\subsubsection{Construction of FCG}
We utilize the \textit{Solidity Compiler}~\cite{solidity} to generate the contract's abstract syntax tree (AST). By analyzing each function node in the AST, we can identify the call relationships for each function. Additionally, modifiers depend on, and visibility attributes associated with each function node are recorded. This step allows us to construct a base function call graph (FCG), where each node includes only the function name, type, and visibility.

Next, we use \textit{Slither}~\cite{slither} to analyze the contract and generate the function-level control flow graph (CFG), represented in SlithIR. These CFGs are then added to their respective function nodes in the FCG. Furthermore, based on the results from the sensitive function localization step, we annotate each function node with a sensitivity label indicating whether it is a sensitive function.

Finally, the enriched FCG captures all the function call relationships between functions and incorporates five types of detailed information for each function node: the function name, type, visibility, sensitivity label, and function-level CFG.

\subsubsection{Access Control Localization}
Sensitive functions are the primary focus of our analysis. Accordingly, we concentrate on the sensitive function nodes within the FCG. Our analysis begins with an access control localization. The execution permissions of functions are typically enforced by verifying the identity of the contract caller, represented in \textit{Solidity} by the special variable $msg.sender$. To identify such permission checks, we search for logical conditions involving $msg.sender$ (e.g., $TMP_1(boolean)=msg.sender==TMP_0$) within the function's CFG. However, some functions may retrieve the value of $msg.sender$ indirectly, either by calling other functions or through logical checks that depend on its value. To address all possible scenarios, we perform data flow analysis on both operands of all logical operators. Suppose either operand is found to be equal to or dependent on $msg.sender$, we consider the function to include a permission check.

However, permission checks are not always performed in the function itself. In many cases, these checks are implemented in the function's modifier or through specialized permission verification functions. To ensure comprehensive coverage, we evaluate three components in our access control search: the \textbf{function itself}, any \textbf{functions it invokes}, and its \textbf{modifiers}. The CFGs for these components are derived directly from the constructed FCG. If a permission check is identified in any of these components, the function is considered to have proper access control.

\subsubsection{Risky Actions Identification}
\label{subsubsec:riskyactions}
\begin{figure}[h]
\centering
\includegraphics[width=\linewidth]{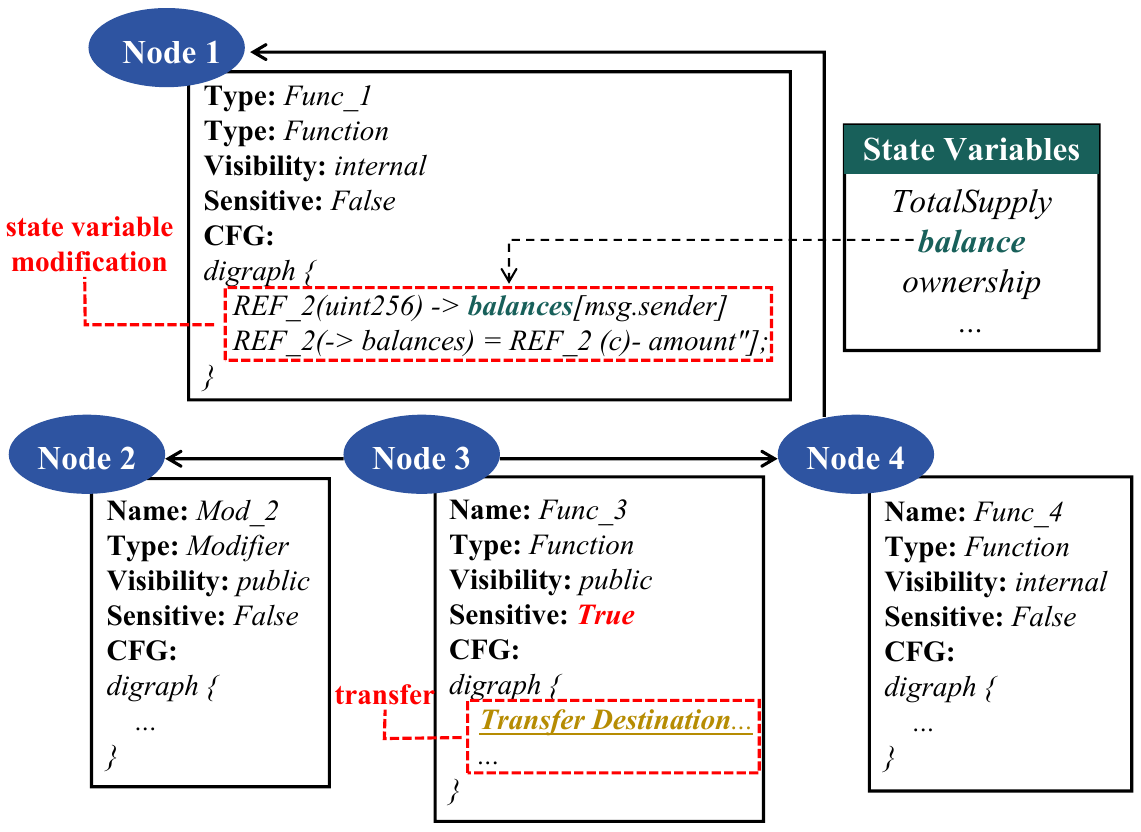}
\caption{The process of searching for \textit{Risky Transfer} in FCG.}
\label{fig:fcg}
\end{figure}

Sensitive functions involve operations that, while potentially leading to vulnerabilities (as shown in Table~\ref{tab:sens_ops}), can sometimes be executed without access control. For example, deposit and withdraw functions may allow anyone to modify state variables and transfer directly. To minimize false positives, we apply stricter criteria to sensitive operations and focus on searching for them:

\noindent \textbf{Risky Transfer.} In smart contracts, fund transfers should typically be accompanied by state modifications to ensure data consistency. For example, a common pattern involves updating a state variable, such as an account balance, before an external call executes the actual transfer. This sequence reflects a standard, safe process. However, a function performing transfers without any associated state variable modifications is considered potentially risky. Figure~\ref{fig:fcg} illustrates our process for identifying such risky actions within a contract's FCG. We begin by examining the sensitive function's CFG for \textit{transfer} operations. If a \textit{transfer} is detected, we then analyze the function and all its callee functions (traversed via the FCG) to ascertain if any state variables (extracted using \textit{Slither}~\cite{slither}) are modified. If a transfer occurs without any corresponding state variable modification within this scope, the function is classified as exhibiting a \textit{Risky Transfer}. Conversely, if both a \textit{transfer} and a state variable modification are found along the execution path, no \textit{Risky Transfer} is flagged.

\noindent \textbf{Risky State Variable Modification.} Similar to \textit{Risky Transfer}, functions that only modify state variables without accompanying transfers can also pose risks, as they may disrupt the normal functioning of the entire contract. For instance, in the $withdraw$ function from Code~\ref{lst:normal_contract}, if only the balance is updated without executing a transfer, the user's balance could decrease without any actual transaction taking place. The detection method for this type of risk is analogous to the approach used for identifying \textit{Risky Transfer}. We begin by searching for operations that modify state variables within the function. Subsequently, we analyze the functions along the call chain in the FCG to identify any transfer operations. If state variable modifications are present without any corresponding transfer operations, the function is deemed to exhibit \textit{Risky State Variable Modification}.

\noindent \textbf{Low-level External Contract Call.} External calls in smart contracts can be made using various methods, including low-level, intermediate, and high-level calls. Low-level calls refer to direct calls to other contracts using the call function, which allows invoking any function at any address and requires the caller to handle the results and errors manually. Among these, low-level calls are the riskiest, as they bypass \textit{Solidity}’s type-checking and error-handling mechanisms. This exposes the contract to potential attacks, where an attacker could design a malicious contract and use a low-level call to trigger its harmful logic within the victim contract. Slither's intermediate representation (SlithIR)~\cite{slithir} clearly distinguishes between different function call methods, with low-level calls denoted as $LOW\_LEVEL\_CALL$. Accordingly, we search the function's CFG for this representation. If it is identified, we classify the function as exhibiting \textit{Low-level External Contract Call}.

\begin{lstlisting}[caption={A smart contract where \textit{Risky State Variable Modification} exists.},label={lst:normal_contract}]
// SPDX-License-Identifier: MIT
pragma solidity ^0.8.0;
contract SimpleBank {
    mapping(address => uint256) private balances;
    function withdraw(uint256 amount) external {
        require(balances[msg.sender] >= amount, "Insufficient balance");
        balances[msg.sender] -= amount;
        // No transfer
    }
}
\end{lstlisting}

\noindent \textbf{Selfdestruct.} 
\textit{Selfdestruct} is a high-risk operation. Once executed, this function irreversibly removes the contract from the blockchain and transfers all remaining balances to a designated address. As a result, it is crucial to implement strict access control mechanisms to prevent unauthorized execution. In SlithIR, the \textit{Selfdestruct} function is represented as $SOLIDITY\_CALL\;selfdestruct$. To identify its presence, we directly search for this representation within the CFG.

While our rule design emphasizes broad coverage of risky patterns, it inevitably introduces the potential for over-approximation in certain scenarios. For example, the \textit{Risky State Variable Modification} rule considers any state-changing operation without an accompanying transfer as potentially risky. However, this assumption may not hold universally—some contracts implement legitimate business logic that updates state without involving any token or value transfer. Recognizing this, \toolname is intentionally designed to report risks rather than confirmed vulnerabilities, offering developers guidance on potentially security-relevant behaviors while leaving final validation to human auditors.

\subsubsection{Detection Algorithm}

The final detection step systematically analyzes the completed smart contract to identify Access Control vulnerabilities. The logic, formalized in Algorithm~\ref{alg:detect_vulnerabilities}, is designed to determine whether sensitive functions contain inadequately protected risky actions. The process begins by constructing a comprehensive FCG of the contract, where each node represents a function and is enriched with its CFG, visibility attributes, and a sensitivity label indicating if it contains sensitive operations.

The algorithm then iterates through the FCG, focusing exclusively on functions marked as sensitive. For each such function, we perform two concurrent analyses. The first, an \textit{Access Control Search}, scans the function's CFG, along with those of its modifiers and any functions it calls, to pinpoint the location of the earliest permission check. Simultaneously, a \textit{Risky Action Search} identifies all instances of the four predefined risky actions (as detailed in Section~\ref{subsubsec:riskyactions}) within the function's CFG.

A vulnerability is ultimately flagged if a risky action is found that is executed before any corresponding access control check, or if no access control check exists at all. This ensures that only actions without prior authorization are considered vulnerabilities. The final output is a comprehensive list of all functions containing such unprotected risky actions, providing developers with precise locations for remediation.

\begin{algorithm}[htbp]
\caption{Detection of Access Control Vulnerabilities}
\label{alg:detect_vulnerabilities}
\begin{algorithmic}[1]
\Require A smart contract \textit{SC}
\Ensure Detected vulnerabilities \textit{VulnList}

\State \textit{FCG} $\leftarrow$ ConstructFCG($SC$)
\State \textit{ACLocation} $\leftarrow$ \textit{null}
\State \textit{RALocations} $\leftarrow \emptyset$
\State \textit{VulnList} $\leftarrow \emptyset$

\For {function $f \in$ \textit{FCG.nodes}}
    \If {\textit{FCG.getLabel}($f$) $\neq$ \textit{sensitive}}
        \State \textbf{continue}
    \EndIf
    \State \textit{ACLocation} $\leftarrow$ AccessControlSearch($f$)
    \State \textit{RALocations} $\leftarrow$ RiskyActionsSearch($f$)

    \If {\textit{RALocations} $\neq \emptyset$}
        \For {location $r \in$ \textit{RALocations}}
            \If {\textit{ACLocation} $= \emptyset$ \textbf{or} \textit{ACLocation} $> r$}
                \State Add $(f, r)$ to \textit{VulnList}
            \EndIf
        \EndFor
    \EndIf
\EndFor

\State \textbf{return} \textit{VulnList}
\end{algorithmic}
\end{algorithm}


\section{Evaluation}
\label{sec:eval}
To evaluate the performance of \toolname, we designed four research questions (RQs) to assess its effectiveness across different stages and scenarios. The RQs are as follows: 

\noindent \textbf{RQ1.} How effective is \toolname in \textit{Sensitive Function Extraction}?

\noindent \textbf{RQ2.} How effective is \toolname in \textit{Function Snippet Completion}?

\noindent \textbf{RQ3.} How does \toolname perform in detecting single smart contracts deployed on blockchain?

\noindent \textbf{RQ4.} How effective is \toolname in detecting real-world smart contract repositories?

\subsection{Experimental Setup}
\subsubsection{Datasets}
\label{subsubsec:datasets}We utilized three datasets for evaluation. The first dataset comprises \textbf{vulnerable smart contracts}, including 15 contracts that have been assigned CVEs~\cite{cve}. These contracts predominantly exhibit typical \ac vulnerabilities and are older versions predating \textit{Solidity} 0.6.0. Notably, this dataset has also been used in prior evaluations of AChecker~\cite{ghaleb2023achecker} and SPCon~\cite{spcon}. The second dataset contains 5,000 recent smart contracts, all of which have been \textbf{deployed on-chain} and verified with validated source code. These contracts were collected from \textit{Etherscan}~\cite{etherscan}, with the latest date being November 12, 2024. The third dataset focuses on \textbf{smart contract repositories} obtained from \textit{DAppSCAN}~\cite{zheng2024dappscan}, which is a large-scale SWC weakness~\cite{swc} dataset from real-world DApps. Using a confidence level of 95\% and a confidence interval of 10~\cite{calculator}, we sampled the \textit{DAppSCAN} dataset and randomly selected 83 smart contract repositories, encompassing a total of 3,092 contracts.


\subsubsection{Setup}
This section details the experimental setup, including the environment and models used to evaluate \toolname's performance.

\paragraph{Experimental environment} Our experiments were conducted on a machine running Ubuntu 22.04.5 LTS, equipped with 10 CPU cores and 20 GB of RAM. To ensure consistency and efficiency, we imposed a time limit of 30 minutes for each tool to analyze a single smart contract.
\paragraph{Model} Apart from using other models for the comparative experiments, we choose to use \gptfouro for all the tasks. All the models were accessed via API calls with default parameter settings.

\subsubsection{Method}
\label{subsubsec:method}
The evaluation dataset comprises both single smart contracts deployed on blockchain and non-compilable smart contract repositories. To accommodate this distinction, we designed two experimental approaches. For single smart contracts deployed on blockchain, all contracts in the dataset are compilable, eliminating the need to address potential compilation failures. In this case, we bypassed intervention with the LLMs by skipping the \textit{Sensitive Function Extraction} and \textit{Function Snippet Completion} steps. Instead, we analyzed every function within the contract as if it were a sensitive function. While this approach may increase the computational cost, it ensures more comprehensive analysis results. For non-compilable smart contract repositories, we applied the methodology detailed in Section~\ref{sec:dappguard} to perform the analysis.

\subsection{Results}
In this section, we present the experimental results addressing the research questions formulated in Section~\ref{sec:eval}. 

\subsubsection{RQ1}
From the 3,092 contracts in the repository dataset, we filtered out secure contracts, such as libraries and interfaces (see Section~\ref{subsec:sens_func_extract}), leaving 1,781 contracts that potentially contain sensitive functions. Using a confidence level of 95\% and a confidence interval of 10, we randomly sampled a subset of 91 contracts from this dataset, following the experimental setup outlined in related research\cite{10.1145/3597926.3598063}. All sensitive functions in the 91 contracts were manually labeled. Specifically, two researchers, each with over two years of experience in smart contract security auditing, independently labeled the sensitive functions. Any disagreements were resolved through discussion to reach a final consensus. This process ensures the quality and consistency of the labels used for evaluation. Subsequently, we tasked the LLM with extracting sensitive functions from the subset. We calculate precision using formula $P=TP/(TP+FP)$, recall using formula $R=TP/(TP+FN)$, F1 using formula $F1=2*P*R/(P+R)$.






Among all the functions analyzed across 91 sampled contracts, the LLM identified 325 true positives (TP), 3 false positives (FP), 913 true negatives (TN), and 10 false negatives (FN). As a result, the LLM achieved an accuracy of 99.1\%, a recall of 97.0\%, and an F1 score of 98.0\% in the task of sensitive function extraction, demonstrating strong performance. Upon analysis of the three false positives, we found that all were caused by hallucinations, where the LLM generated a function signature that did not exist in the original code. Conversely, the majority of the ten false negatives resulted from the extensive length of certain sensitive functions, which negatively impacted the model’s inference performance. 
\vspace{-0.2cm}
\begin{center}
    \begin{myboxc} \textbf{Answer to RQ1:} \toolname demonstrated excellent performance in sensitive function extraction, achieving an F1 score of \textit{98.0\%}. 
    \end{myboxc}
\end{center}

\subsubsection{RQ2}
In the repository dataset, there are 83 projects comprising a total of 3,092 contract files. From these files, we extracted 4,263 sensitive functions, each of which was used to generate a completed smart contract, resulting in 4,263 completed contracts. To evaluate \toolname’s capability in function snippet completion, we employed two metrics: 1) \textbf{Number of contracts that are uncompilable}: Our objective is to compile contracts containing sensitive functions and subsequently conduct static analysis. Therefore, the ability of the completed contracts to successfully compile is a critical criterion. We used the \textit{Solidity} compiler to compile these contracts and recorded the number of failures. 2)	\textbf{Number of contracts where the original code was modified}: In addition to ensuring successful compilation, it is vital that the original sensitive function snippets remain unaltered during the completion process. Modifications to the original code could introduce new vulnerabilities or inadvertently resolve existing ones, potentially affecting the accuracy of the final detection results. To determine whether the original code was modified, we preprocessed the code before and after completion by removing irrelevant characters (e.g., extra spaces, line breaks, comments). We then attempted to locate the original code in the completed contract. If the search failed, we considered the original code to have been modified, and any subsequent results will be disregarded.

According to the experimental results, out of a total of 4,263 contracts, 4,169 were successfully compiled after the process of completion and self-reflection, achieving a success rate of 97.8\%. Furthermore, 4,051 contracts remained unmodified, accounting for 95.0\% of the total. These results demonstrate that the LLM performs effectively in function snippet completion.


\begin{center}
    \begin{myboxc} \textbf{Answer to RQ2:} \toolname achieved a 97.8\% compilation success rate and ensured 95.0\% of contracts being completed correctly, demonstrating strong performance in function snippet completion.
    \end{myboxc}
\end{center}

\subsubsection{RQ3}
We began by evaluating \toolname using a dataset containing known vulnerabilities. Following the experimental framework outlined by \textit{AChecker}~\cite{ghaleb2023achecker}, we included a selection of tools integrated with smartbugs~\cite{FerreiraEtAl2020ASE} (i.e., \textit{Slither}~\cite{slither}, \textit{Maian}~\cite{maian}, \textit{Manticore}~\cite{mossberg2019manticore}, \textit{SmartCheck}~\cite{smartcheck}, and \textit{Mythril}~\cite{mythril}), each capable of detecting at least one type of \ac vulnerability. Additionally, \textit{SPCon}~\cite{spcon} and \textit{GPTScan} were incorporated into this comparative analysis. We also included \textit{AChecker}~\cite{ghaleb2023achecker} itself as a baseline for comparison. Certain tools were excluded from our analysis because they failed to identify any vulnerabilities, reporting no positive results. The experimental results for all tools on this dataset are presented in Table~\ref{tab:cve_recall}.

\begin{table}[ht]
\caption{Comparison with other SOTA tools on the vulnerable dataset.}
\label{tab:cve_recall}
\centering
\begin{tabular}{lccccccc}
\textbf{CVE} & 
\rotatebox{90}{\textbf{Slither}} & 
\rotatebox{90}{\textbf{Maian}} & 
\rotatebox{90}{\textbf{SmartCheck}} & 
\rotatebox{90}{\textbf{Mythril}} & 
\rotatebox{90}{\textbf{SPCon}} & 
\rotatebox{90}{\textbf{AChecker}} &
\rotatebox{90}{\textbf{Ours}}
\\
\midrule
CVE-2018-10666 & & & & & \checkmark& \checkmark&\checkmark\\
CVE-2018-10705 & & & & & \checkmark& \checkmark&\checkmark\\
CVE-2018-11329 & & & & & \checkmark& \checkmark&\checkmark\\
CVE-2018-19830 & & & & & & \checkmark&\checkmark\\
CVE-2018-19831 & & & & \checkmark & \checkmark& \checkmark&\checkmark\\
CVE-2018-19832 & & \checkmark & & \checkmark & \checkmark& \checkmark&\checkmark\\
CVE-2018-19833 & & & & & & \checkmark&\checkmark\\
CVE-2018-19834 & & & & & & \checkmark& \checkmark\\
CVE-2019-15078 & & \checkmark & & \checkmark & \checkmark& \checkmark& \checkmark\\
CVE-2019-15079 & & & & & \checkmark& & \checkmark \\
CVE-2019-15080 & & & & & \checkmark& \checkmark& \checkmark \\
CVE-2020-17753 & \checkmark & & \checkmark & \checkmark & & & \checkmark \\
CVE-2020-35962 & & & & & & & \\
CVE-2021-34272 & & & & & \checkmark& \checkmark& \checkmark\\
CVE-2021-34273 & & & & & & \checkmark& \checkmark\\
\midrule
\textbf{Recall\%} & 6 & 13 & 6 & 26 & 60 & 80 & \textbf{93} \\
\bottomrule
\end{tabular}
\end{table}
\setlength{\textfloatsep}{3pt}
\toolname successfully detected 14 out of 15 CVEs, achieving a recall rate of 93\%. The one missed vulnerability, CVE-2020-35962, could not be detected due to compilation issues; notably, no other tool was able to identify this vulnerability either. Among the comparative tools, \textit{AChecker} performed the best, identifying 12 CVEs and achieving a recall rate of 80\%. These results underscore the superior comprehensiveness of \toolname in detecting vulnerabilities in smart contracts.

Code~\ref{lst:comparison} presents a simplified version of CVE-2019-15079, an access control vulnerability that evaded detection by \textit{AChecker} but was successfully identified by \toolname. The vulnerability in this contract arises from a \textbf{misnamed constructor}, i.e., \texttt{EAI\_TokenERC} was misnamed as \texttt{EAI\_TokenERC20}, making a sensitive initialization function publicly accessible and callable by any user. \toolname detects this issue by leveraging an LLM to recognize the function's privileged semantics (e.g., state variable modification) and identifies the lack of access control through static analysis. In contrast, \textit{AChecker} missed this vulnerability because it failed to recognize the function as sensitive, due to its reliance on predefined patterns for critical instructions. Moreover, it lacks semantic reasoning capabilities to detect misnamed constructors or to infer intended behavior from code context. As a result, it cannot flag the public accessibility of a privileged initialization function as a security risk.

\begin{lstlisting}[caption={Vulnerabilities that cannot be detected by AChecker but can be detected by Trace.},label={lst:comparison}]
pragma solidity ^0.4.16;
contract EAI_TokenERC {
    uint256 public totalSupply;
    mapping (address => uint256) public balanceOf;    
    function EAI_TokenERC20(
        uint256 initialSupply,
        string tokenName,
        string tokenSymbol
    ) public {
        totalSupply = initialSupply * 10 ** 8;
        balanceOf[msg.sender] = totalSupply;
    }
}
\end{lstlisting}

Since the contracts in the CVE dataset all contain vulnerabilities and the size of the dataset is limited, we also validate the effectiveness of \toolname on a larger-scale real-world dataset (on-chain dataset). For the experiments, we selected the same tools used for comparison. Notably, some tools are limited to accurately analyzing early versions of \textit{Solidity}~\cite{solidity} (e.g., versions prior to v0.6.0), whereas most contracts in the on-chain dataset are written using newer \textit{Solidity} versions (e.g., v0.8.0 and later). As a result, tools like \textit{Oyente}~\cite{oyente} and \textit{SmartCheck}~\cite{smartcheck} were excluded from the comparison. Furthermore, \textit{SPCon}~\cite{spcon}, which relies on the transaction history of smart contracts for analysis, could not be applied because the latest on-chain contracts have insufficient transaction records to meet its analysis requirements. The final experimental results are presented in Table~\ref{tab:onchain}.

\begin{table}[!t]
\caption{Comparison with SOTA tools on the on-chain dataset.\label{tab:onchain}}
\centering
\renewcommand{\arraystretch}{1.1} 
\small
\begin{tabular}{ccccc}
\hline
\textbf{Tools} & \textbf{\# TP} & \textbf{\# FP} & \textbf{\# Failure} & \textbf{Precision\%}\\
\hline
\textit{Manticore} & 0 & 0 & 322 & 0 \\
\textit{Maian} & 0 & 0 & 1,447 & 0 \\
\textit{Mythril} & 1 & 2 & 601 & 33.3 \\
\textit{AChecker} & 4 & 7 & 485 & 36.3 \\
\textit{Slither} & 90 & 71 & 35 & 55.9 \\
\textit{GPTScan} & 80 & 24 & 76 & 76.9 \\
\hline
\toolname & 91 & 11 & 58 & \textbf{89.2} \\
\hline
\end{tabular}
\end{table}

We recorded all positive results reported by the tools and manually verified each one. Each positive result was manually verified by two smart contract researchers, each with over two years of experience. Discrepancies in their findings were resolved through discussion until a consensus was reached. This entire verification process concluded within two weeks. The analysis revealed that \toolname detected 91 true positives and 11 false positives, achieving a precision of 89.2\%. This result makes \toolname the most precise tool, outperforming the second-ranked \textit{GPTScan}~\cite{10.1145/3597503.3639117} by a margin of 12.3\%. Robustness is another critical factor in evaluating tool performance, as some tools encounter timeouts when analyzing highly complex contracts. Thus, beyond positive results, we also considered the number of contracts that each tool failed to analyze. As shown in Table~\ref{tab:onchain}, \textit{Slither} had the lowest number of failed analyses, with 35 contracts, while \toolname reported slightly more failures, at 58. This is partly due to \toolname’s approach of analyzing each function independently when evaluating individual contracts (see Section~\ref{subsubsec:method}), which increases computational costs. However, as the results demonstrate, this trade-off remains acceptable, given the tool’s superior detection performance.

\begin{center}
    \begin{myboxc} \textbf{Answer to RQ3:} \toolname demonstrated superior detection performance, achieving a 93\% recall on CVEs and 89.2\% precision on a large-scale on-chain dataset, outperforming comparative SOTA tools.
    \end{myboxc}
\end{center}

\subsubsection{RQ4}
\label{subsubsec:rq4}


We compared \toolname with a baseline LLM for detecting vulnerabilities in non-compilable repositories. Recognizing that strict output formatting constraints can negatively affect LLM performance and increase hallucination rates—as observed in prior work~\cite{chen2023chatgpt}—we followed their recommendation. A two-stage prompting strategy was adopted to mitigate this issue.

\begin{figure}[ht]
\centering
\includegraphics[width=\linewidth]{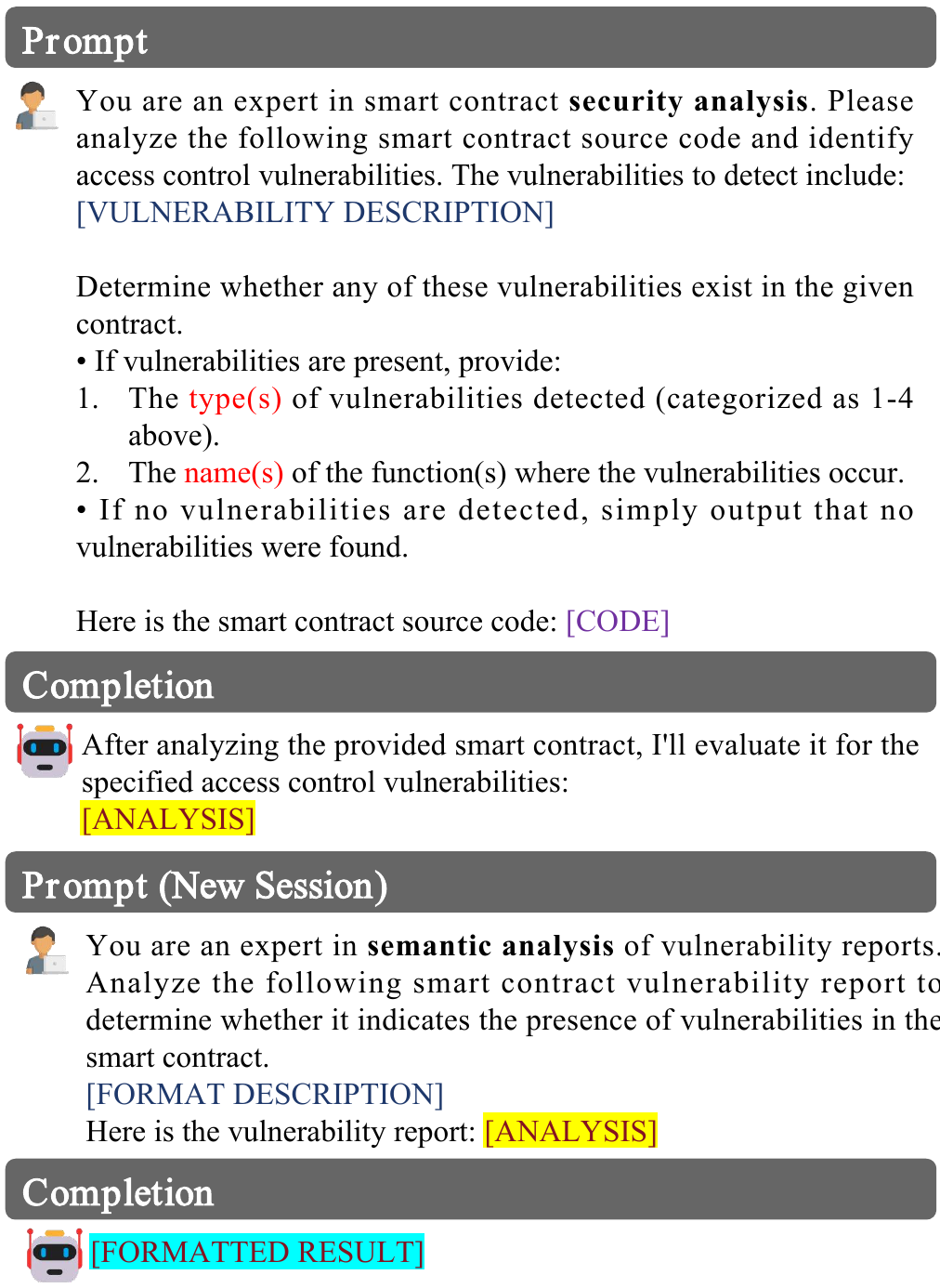}
\caption{Two-stage prompt template for vulnerability detection using LLMs. The first prompt performs free-form analysis on the contract source, while the second prompt parses and formats the analysis result.}
\label{fig:llm_baseline}
\end{figure}

As illustrated in Fig.~\ref{fig:llm_baseline}, the first stage instructs the LLM to act as a smart contract security expert and perform unconstrained vulnerability analysis on a given contract. It receives as input the source code along with a description of the four types of access control vulnerabilities. It outputs a natural-language analysis that either (1) lists the type(s) of vulnerabilities detected and the name(s) of the affected functions, or (2) states that no vulnerabilities were found. In the second stage, a new LLM session is initiated to semantically parse the output from the first stage and convert it into a structured format compatible with downstream processing. This separation between reasoning and formatting improves reliability by avoiding complex prompt constraints during analysis, while still producing consistent outputs for evaluation.

\begin{table}[!t]
\caption{Comparison with LLMs on the repository dataset.\label{tab:dapp}}
\centering
\renewcommand{\arraystretch}{1.1} 
\small
\begin{tabular}{ccccc}
\hline
\textbf{LLMs} & \textbf{\# TP} & \textbf{\# FP} & \textbf{\# Failure} & \textbf{Precision\%}\\
\hline
Gemini-2-flash-exp & 34 & 547 & 0 & 5.6\\
Claude-3.5-sonnet & 28 & 387 & 0 & 6.7\\
o1 & 14 & 108 & 0 & 11.5\\
GPT-4o & 22 & 141 & 0 & 13.5 \\
DeepSeek-R1 & 26 & 156 & 0 & 14.3 \\
\hline
\toolname & 40 & 6 & 5 & \textbf{87.0} \\
\hline
\end{tabular}
\end{table}

We selected five of the latest and widely used models for comparison: \textit{Gemini-2-flash-exp}~\cite{gemini}, \textit{Claude-3.5-sonnet}~\cite{claude3.5}, \textit{o1}~\cite{o1}, \textit{GPT-4o}~\cite{gpt4o} and \textit{DeepSeek-R1}~\cite{deepseekai2025deepseekr1incentivizingreasoningcapability}. The experimental results are summarized in Table~\ref{tab:dapp}. Among these models, \textit{DeepSeek-R1} achieved the highest precision, detecting 26 true positives (TPs) with a precision of 14.3\%. However, it also produced a significant number of false positives, totaling 156. This aligns with findings from prior research~\cite{chen2023chatgpt}, which highlight that LLMs often generate a high volume of false positives during vulnerability detection tasks, leading to low overall precision.

In contrast, \toolname does not rely on LLMs for direct vulnerability detection. Instead, it utilizes LLMs exclusively for code understanding and completion. The actual vulnerability detection process is carried out through static analysis, effectively mitigating the issue of excessive false positives associated with LLM-based detection. As a result, \toolname identified 40 TPs with a substantially higher precision of 87.0\%. Nevertheless, LLMs demonstrate superior robustness. All five models successfully processed the contracts without encountering detection failures, thanks to their ability to support extended input lengths. In contrast, \toolname failed to analyze five contracts due to the complexity of the completed contracts, leading to analysis timeouts. However, given \toolname’s significantly higher precision, we consider these occasional timeouts to be an acceptable trade-off.

Our analysis of \toolname's six false positives showed that four were caused by the LLM component, which introduced new vulnerabilities by modifying secure code. The other two were a direct result of a design trade-off: to manage computational costs, our static analyzer is limited to a function call depth of three. Consequently, it failed to detect access control statements located in call chains deeper than this pre-set limit.
\begin{center}
    \begin{myboxc} \textbf{Answer to RQ4:} \toolname achieved a precision of 87.0\% on smart contract repositories, outperforming five popular LLMs.
    \end{myboxc}
\end{center}
\section{Discussion}
\subsection{Implications}
\noindent \textbf{For Third-Party Developers.} \toolname directly addresses the pain points of code reuse and security validation in decentralized development. By removing the need for compilation, the tool broadens access to security analysis, enabling developers to efficiently analyze complex, dependency-laden repositories.

\noindent \textbf{For Smart Contract Repository Development Teams.} While development teams typically operate within compilable environments, \toolname offers a competitive advantage in security assurance. Its superior detection performance compared to SOTA tools (see Section~\ref{sec:eval}) suggests that even teams with mature workflows could benefit from integrating \toolname into their security pipelines.

\noindent \textbf{For Researchers.} Our approach allows for large-scale analysis of historical repositories that are non-compilable due to deprecated dependencies or toolchain changes, enabling in-depth exploration of their security aspects. Moreover, the tool’s architecture is extendable, allowing for the integration of additional detection heuristics and the incorporation of new vulnerability types, thereby supporting ongoing advancements in smart contract security research.



\subsection{Complexity of Compiling Repositories}
\label{complexity}
We performed local compilation on the repository-level dataset, which contains 83 smart contract repositories. This included attempts to compile both single smart contract files in the repository and the entire repository. Table~\ref{tab:complexity} presents key complexity metrics related to the dataset. At the file level, we used the exact solc version specified in each of the 3,092 contracts. Widespread dependency and compatibility issues resulted in a low compilation success rate of only 30.6\%. 

At the repository level, we used Truffle~\cite{Truffle}\footnote{\textit{Truffle} has been officially deprecated. We use it here for historical consistency with some older repositories in the dataset.} to compile all 83 projects. Compilation was further complicated by dependencies on 13 external libraries, four of which lack Truffle support. This led to an even lower success rate of 6.0\% (5/83), highlighting that the vast majority of repositories cannot be compiled without significant manual intervention.

\begin{table}[!t]
\caption{Complexity of repository dataset.\label{tab:complexity}}
\centering
\renewcommand{\arraystretch}{1.15} 
\small
\begin{tabular}{p{3.8cm}p{1.2cm}p{2.6cm}}
\hline
\textbf{Metric} & \textbf{Level} & \textbf{Value}\\
\hline
Average lines of code                             & Contract & 139.4                                \\ 
Internal file dependencies                        & Contract & 6,792                                \\ 
Compilation success rate        & Contract &  945 / 3,092 (30.6\%)                  \\ 
\hline
Average number of contracts         & Repository & 37.3                                 \\ 
External library dependencies            & Repository         & 13 (4 unsupported) \\ 
Compilation success rate    & Repository        & 5 / 83 (6.0\%)                         \\ 
\hline
\end{tabular}
\end{table}

\subsection{Threats to Validity}
\label{subsec:threats}
\textbf{Internal Threats.} 
The use of LLMs introduces inherent time and computational costs. However, this overhead is justified by the ability of LLMs to fully automate the analysis of smart contract repositories -- an advantage over traditional tools that often require manual effort to ensure successful compilation. While this work focuses specifically on access control vulnerabilities, the underlying framework of \toolname is extensible. It can be adapted to support the detection of additional vulnerability types by extending the set of sensitive operations and analysis rules accordingly.

\textbf{External Threats.} During the sensitive function extraction and function snippet completion stages, \toolname leverages LLMs, which are prone to inherent limitations such as hallucination and data leakage. To address this, we validate the results generated by the LLMs. If the results contain non-existent function signatures or alter the original code, \toolname will detect and report these discrepancies. 
\section{Related Work}
\label{sec:related}
\subsection{Large Language Models}
Large Language Models (LLMs) have been widely applied in various tasks in the field of software engineering. Ahmed et al.~\cite{10.1145/3551349.3559555} investigate the use of few-shot training with the GPT Codex model and demonstrate that project-specific training can significantly outperform state-of-the-art models for code summarization. Wang et al.~\cite{wang2024rlcoderreinforcementlearningrepositorylevel} propose \textit{RLCoder}, a novel reinforcement learning framework for repository-level code completion that enables unsupervised learning of useful content retrieval. Chen et al.~\cite{chen2023chatgpt} and David et al.~\cite{david2023needmanualsmartcontract} studied the performance of LLMs in detecting smart contract vulnerabilities and identified their low precision. Unlike these approaches, we do not directly employ LLMs for vulnerability detection. Instead, we leverage them for code understanding and completion to support program analysis. Kim et al.~\cite{10830749} explore the use of a lightweight transformer-based NLP model to detect vulnerabilities in smart contracts. Ma et al.~\cite{ma2024combining} present \textit{iAudit}, a multi-agent system for detecting and explaining Solidity smart-contract vulnerabilities.

\subsection{Smart Contract Vulnerability Detection}
Smart contract vulnerability detection has also received widespread attention from researchers. Static analysis is a widely used method for detecting vulnerabilities in smart contracts, as it examines the source code without executing it, allowing for the identification of potential issues early in the development process~\cite{1366126}. Luu et al.~\cite{10.1145/2976749.2978309} develop a symbolic execution tool, \textit{Oyente}, which identifies vulnerabilities in existing contracts. Chen et al.~\cite{9072659} identify 20 types of defects in smart contracts, and provide a dataset to help developers prioritize defect removal for improved contract quality. Feist et al.~\cite{slither} present \textit{Slither}, a static analysis framework for Ethereum smart contracts that converts \textit{Solidity} contracts into an intermediate representation for efficient bug detection, optimization, and code review. Tikhomirov et al.~\cite{10.1145/3194113.3194115} provide a comprehensive classification of Solidity code issues and present \textit{SmartCheck}, an extensible static analysis tool for detecting vulnerabilities in smart contracts. Bose et al.~\cite{9833721} present \textit{Sailfish}, a scalable system for detecting state-inconsistency bugs in smart contracts, combining a lightweight exploration phase with precise symbolic evaluation guided by value-summary analysis. Dynamic analysis is also an effective solution for vulnerability detection~\cite{10.5555/318773.318944}. Mossberg et al. introduce \textit{Manticore}~\cite{8952204}, a dynamic symbolic execution framework for analyzing binaries and Ethereum smart contracts. Grieco et al.~\cite{10.1145/3395363.3404366} introduce \textit{Echidna}, a smart contract fuzzer designed for automatic test generation to detect property violations. Traditional programming analysis methods can effectively detect vulnerabilities in smart contracts, but they often rely on the assumption that the contract is compilable. In contrast, \toolname leverages LLMs to overcome this limitation, enabling vulnerability detection in non-compilable smart contract repositories.
\section{Conclusion}
\label{sec:conclusion}

We introduced \toolname, a novel tool for detecting vulnerabilities in non-compilable smart contract repositories by combining LLM capabilities with traditional static analysis. Evaluations show \toolname robustly outperforms existing tools: it achieved 89.2\% precision on an on-chain dataset, exceeding \textit{GPTScan} by 12.3\%, and 87.0\% precision on a repository dataset, significantly surpassing \textit{DeepSeek-R1} (14.3\%). This work underscores the promise of combining LLMs with program analysis for intricate smart contract security challenges.

\bibliographystyle{IEEEtran}
\bibliography{refs}

\begin{thebibliography}{10}
\providecommand{\url}[1]{#1}
\csname url@samestyle\endcsname
\providecommand{\newblock}{\relax}
\providecommand{\bibinfo}[2]{#2}
\providecommand{\BIBentrySTDinterwordspacing}{\spaceskip=0pt\relax}
\providecommand{\BIBentryALTinterwordstretchfactor}{4}
\providecommand{\BIBentryALTinterwordspacing}{\spaceskip=\fontdimen2\font plus
\BIBentryALTinterwordstretchfactor\fontdimen3\font minus \fontdimen4\font\relax}
\providecommand{\BIBforeignlanguage}[2]{{%
\expandafter\ifx\csname l@#1\endcsname\relax
\typeout{** WARNING: IEEEtran.bst: No hyphenation pattern has been}%
\typeout{** loaded for the language `#1'. Using the pattern for}%
\typeout{** the default language instead.}%
\else
\language=\csname l@#1\endcsname
\fi
#2}}
\providecommand{\BIBdecl}{\relax}
\BIBdecl

\bibitem{chen2021defectchecker}
J.~Chen, X.~Xia, D.~Lo, J.~Grundy, X.~Luo, and T.~Chen, ``Defectchecker: Automated smart contract defect detection by analyzing evm bytecode,'' \emph{IEEE Transactions on Software Engineering}, vol.~48, no.~7, pp. 2189--2207, 2021.

\bibitem{cai2018decentralized}
W.~Cai, Z.~Wang, J.~B. Ernst, Z.~Hong, C.~Feng, and V.~C. Leung, ``Decentralized applications: The blockchain-empowered software system,'' \emph{IEEE access}, vol.~6, pp. 53\,019--53\,033, 2018.

\bibitem{chitta2019decentralized}
S.~Chitta, S.~M. Yellepeddi, S.~Thota, and A.~K.~P. Venkata, ``Decentralized finance (defi): A comprehensive study of protocols and applications,'' \emph{Distributed Learning and Broad Applications in Scientific Research}, vol.~5, pp. 124--145, 2019.

\bibitem{muller2019hidals}
M.~M{\"u}ller, S.~R. Garzon, M.~Westerkamp, and Z.~A. Lux, ``Hidals: A hybrid iot-based decentralized application for logistics and supply chain management,'' in \emph{2019 IEEE 10th Annual Information Technology, Electronics and Mobile Communication Conference (IEMCON)}.\hskip 1em plus 0.5em minus 0.4em\relax IEEE, 2019, pp. 0802--0808.

\bibitem{satybaldy2022decentralized}
A.~Satybaldy, A.~Hasselgren, and M.~Nowostawski, ``Decentralized identity management for e-health applications: State-of-the-art and guidance for future work,'' \emph{Blockchain in Healthcare Today}, vol.~5, 2022.

\bibitem{ac}
\BIBentryALTinterwordspacing
OWASP, ``Vulnerability: Improper access control.'' [Online]. Available: \url{https://owasp.org/www-project-smart-contract-top-10/2023/en/src/SC04-access-control-vulnerabilities.html}
\BIBentrySTDinterwordspacing

\bibitem{paritybugs}
\BIBentryALTinterwordspacing
L.~Breidenbach, P.~Daian, A.~Juels, and E.~G. Sirer, ``An in-depth look at the parity multisig bug,'' 2017. [Online]. Available: \url{https://hackingdistributed.com/2017/07/22/deep-dive-parity-bug/}
\BIBentrySTDinterwordspacing

\bibitem{slither}
J.~Feist, G.~Grieco, and A.~Groce, ``Slither: a static analysis framework for smart contracts,'' in \emph{2019 IEEE/ACM 2nd International Workshop on Emerging Trends in Software Engineering for Blockchain (WETSEB)}.\hskip 1em plus 0.5em minus 0.4em\relax IEEE, 2019, pp. 8--15.

\bibitem{oyente}
L.~Luu, D.-H. Chu, H.~Olickel, P.~Saxena, and A.~Hobor, ``Making smart contracts smarter,'' in \emph{Proceedings of the 2016 ACM SIGSAC conference on computer and communications security}, 2016, pp. 254--269.

\bibitem{smartcheck}
S.~Tikhomirov, E.~Voskresenskaya, I.~Ivanitskiy, R.~Takhaviev, E.~Marchenko, and Y.~Alexandrov, ``Smartcheck: Static analysis of ethereum smart contracts,'' in \emph{Proceedings of the 1st international workshop on emerging trends in software engineering for blockchain}, 2018, pp. 9--16.

\bibitem{securify}
P.~Tsankov, A.~Dan, D.~Drachsler-Cohen, A.~Gervais, F.~Buenzli, and M.~Vechev, ``Securify: Practical security analysis of smart contracts,'' in \emph{Proceedings of the 2018 ACM SIGSAC conference on computer and communications security}, 2018, pp. 67--82.

\bibitem{mossberg2019manticore}
M.~Mossberg, F.~Manzano, E.~Hennenfent, A.~Groce, G.~Grieco, J.~Feist, T.~Brunson, and A.~Dinaburg, ``Manticore: A user-friendly symbolic execution framework for binaries and smart contracts,'' in \emph{2019 34th {IEEE/ACM} International Conference on Automated Software Engineering ({ASE})}, 2019, pp. 1186--1189.

\bibitem{solidity}
\BIBentryALTinterwordspacing
``Solidity,'' 2024. [Online]. Available: \url{https://soliditylang.org}
\BIBentrySTDinterwordspacing

\bibitem{10.1145/3597503.3639117}
\BIBentryALTinterwordspacing
Y.~Sun, D.~Wu, Y.~Xue, H.~Liu, H.~Wang, Z.~Xu, X.~Xie, and Y.~Liu, ``Gptscan: Detecting logic vulnerabilities in smart contracts by combining gpt with program analysis,'' in \emph{Proceedings of the IEEE/ACM 46th International Conference on Software Engineering}, ser. ICSE '24.\hskip 1em plus 0.5em minus 0.4em\relax New York, NY, USA: Association for Computing Machinery, 2024. [Online]. Available: \url{https://doi.org/10.1145/3597503.3639117}
\BIBentrySTDinterwordspacing

\bibitem{ghaleb2023achecker}
A.~Ghaleb, J.~Rubin, and K.~Pattabiraman, ``Achecker: Statically detecting smart contract access control vulnerabilities,'' in \emph{2023 IEEE/ACM 45th International Conference on Software Engineering (ICSE)}.\hskip 1em plus 0.5em minus 0.4em\relax IEEE, 2023, pp. 945--956.

\bibitem{spcon}
Y.~Liu, Y.~Li, S.-W. Lin, and C.~Artho, ``Finding permission bugs in smart contracts with role mining,'' in \emph{Proceedings of the 31st ACM SIGSOFT International Symposium on Software Testing and Analysis}, 2022, pp. 716--727.

\bibitem{ethereum}
\BIBentryALTinterwordspacing
``Ethereum,'' 2024. [Online]. Available: \url{https://ethereum.org/en/}
\BIBentrySTDinterwordspacing

\bibitem{mythril}
\BIBentryALTinterwordspacing
ConsenSys, ``Mythril: Security analysis tool for evm bytecode,'' 2021. [Online]. Available: \url{https://github.com/ConsenSys/mythril}
\BIBentrySTDinterwordspacing

\bibitem{maian}
I.~Nikoli{\'c}, A.~Kolluri, I.~Sergey, P.~Saxena, and A.~Hobor, ``Finding the greedy, prodigal, and suicidal contracts at scale,'' in \emph{Proceedings of the 34th annual computer security applications conference}, 2018, pp. 653--663.

\bibitem{zheng2024dappscan}
Z.~Zheng, J.~Su, J.~Chen, D.~Lo, Z.~Zhong, and M.~Ye, ``Dappscan: building large-scale datasets for smart contract weaknesses in dapp projects,'' \emph{IEEE Transactions on Software Engineering}, 2024.

\bibitem{Hardhat}
\BIBentryALTinterwordspacing
``Hardhat,'' 2024. [Online]. Available: \url{https://hardhat.org/}
\BIBentrySTDinterwordspacing

\bibitem{Truffle}
\BIBentryALTinterwordspacing
``Truffle suite,'' 2024. [Online]. Available: \url{https://archive.trufflesuite.com/}
\BIBentrySTDinterwordspacing

\bibitem{openzeppelin}
\BIBentryALTinterwordspacing
``Openzeppelin,'' 2024. [Online]. Available: \url{https://www.openzeppelin.com}
\BIBentrySTDinterwordspacing

\bibitem{staticanalysis}
N.~Ayewah, W.~Pugh, D.~Hovemeyer, J.~D. Morgenthaler, and J.~Penix, ``Using static analysis to find bugs,'' \emph{IEEE Software}, vol.~25, no.~5, pp. 22--29, 2008.

\bibitem{evm}
\BIBentryALTinterwordspacing
Wikipedia, ``Ethereum virtual machine,'' 2024. [Online]. Available: \url{https://en.wikipedia.org/wiki/Ethereum#Virtual_machine}
\BIBentrySTDinterwordspacing

\bibitem{vyper}
\BIBentryALTinterwordspacing
``Vyper,'' 2024. [Online]. Available: \url{https://docs.vyperlang.org/en/stable/}
\BIBentrySTDinterwordspacing

\bibitem{chen2024identifying}
J.~Chen, C.~Chen, J.~Hu, J.~Grundy, Y.~Wang, T.~Chen, and Z.~Zheng, ``Identifying smart contract security issues in code snippets from stack overflow,'' in \emph{Proceedings of the 33rd ACM SIGSOFT International Symposium on Software Testing and Analysis}, 2024, pp. 1198--1210.

\bibitem{slithir}
\BIBentryALTinterwordspacing
crytic, ``Slithir,'' 2024. [Online]. Available: \url{https://github.com/crytic/slither/wiki/SlithIR}
\BIBentrySTDinterwordspacing

\bibitem{cve}
\BIBentryALTinterwordspacing
``Cve,'' 2024. [Online]. Available: \url{https://www.cve.org}
\BIBentrySTDinterwordspacing

\bibitem{etherscan}
\BIBentryALTinterwordspacing
``Etherscan,'' 2024. [Online]. Available: \url{https://etherscan.io}
\BIBentrySTDinterwordspacing

\bibitem{swc}
\BIBentryALTinterwordspacing
``Smart contract weakness classification,'' 2024. [Online]. Available: \url{https://swcregistry.io/}
\BIBentrySTDinterwordspacing

\bibitem{calculator}
\BIBentryALTinterwordspacing
``Sample size calculator,'' 2024. [Online]. Available: \url{https://www.surveysystem.com/sscalc.htm}
\BIBentrySTDinterwordspacing

\bibitem{10.1145/3597926.3598063}
\BIBentryALTinterwordspacing
S.~Yang, J.~Chen, and Z.~Zheng, ``Definition and detection of defects in nft smart contracts,'' in \emph{Proceedings of the 32nd ACM SIGSOFT International Symposium on Software Testing and Analysis}, ser. ISSTA 2023.\hskip 1em plus 0.5em minus 0.4em\relax New York, NY, USA: Association for Computing Machinery, 2023, p. 373–384. [Online]. Available: \url{https://doi.org/10.1145/3597926.3598063}
\BIBentrySTDinterwordspacing

\bibitem{FerreiraEtAl2020ASE}
J.~F. Ferreira, P.~Cruz, T.~Durieux, and R.~Abreu, ``{SmartBugs}: A framework to analyze {Solidity} smart contracts,'' in \emph{Proceedings of the 35th IEEE/ACM International Conference on Automated Software Engineering}, 2020, pp. 1349--1352.

\bibitem{chen2023chatgpt}
C.~Chen, J.~Su, J.~Chen, Y.~Wang, T.~Bi, J.~Yu, Y.~Wang, X.~Lin, T.~Chen, and Z.~Zheng, ``When chatgpt meets smart contract vulnerability detection: How far are we?'' \emph{ACM Transactions on Software Engineering and Methodology}, 2023.

\bibitem{gemini}
\BIBentryALTinterwordspacing
Google, ``Gemini 2.0 flash,'' 2024. [Online]. Available: \url{https://ai.google.dev/gemini-api/docs/models/gemini-v2?hl=zh-cn}
\BIBentrySTDinterwordspacing

\bibitem{claude3.5}
\BIBentryALTinterwordspacing
Anthropic, ``Claude 3.5 sonnet,'' 2024. [Online]. Available: \url{https://www.anthropic.com/claude/sonnet}
\BIBentrySTDinterwordspacing

\bibitem{o1}
\BIBentryALTinterwordspacing
OpenAI, ``o1,'' 2024. [Online]. Available: \url{https://openai.com/o1/}
\BIBentrySTDinterwordspacing

\bibitem{gpt4o}
\BIBentryALTinterwordspacing
------, ``Gpt-4o,'' 2024. [Online]. Available: \url{https://platform.openai.com/docs/models#gpt-4o}
\BIBentrySTDinterwordspacing

\bibitem{deepseekai2025deepseekr1incentivizingreasoningcapability}
\BIBentryALTinterwordspacing
DeepSeek-AI, D.~Guo, D.~Yang, H.~Zhang, J.~Song, and et~al., ``Deepseek-r1: Incentivizing reasoning capability in llms via reinforcement learning,'' 2025. [Online]. Available: \url{https://arxiv.org/abs/2501.12948}
\BIBentrySTDinterwordspacing

\bibitem{10.1145/3551349.3559555}
\BIBentryALTinterwordspacing
T.~Ahmed and P.~Devanbu, ``Few-shot training llms for project-specific code-summarization,'' in \emph{Proceedings of the 37th IEEE/ACM International Conference on Automated Software Engineering}, ser. ASE '22.\hskip 1em plus 0.5em minus 0.4em\relax New York, NY, USA: Association for Computing Machinery, 2023. [Online]. Available: \url{https://doi.org/10.1145/3551349.3559555}
\BIBentrySTDinterwordspacing

\bibitem{wang2024rlcoderreinforcementlearningrepositorylevel}
\BIBentryALTinterwordspacing
Y.~Wang, Y.~Wang, D.~Guo, J.~Chen, R.~Zhang, Y.~Ma, and Z.~Zheng, ``Rlcoder: Reinforcement learning for repository-level code completion,'' 2024. [Online]. Available: \url{https://arxiv.org/abs/2407.19487}
\BIBentrySTDinterwordspacing

\bibitem{david2023needmanualsmartcontract}
\BIBentryALTinterwordspacing
I.~David, L.~Zhou, K.~Qin, D.~Song, L.~Cavallaro, and A.~Gervais, ``Do you still need a manual smart contract audit?'' 2023. [Online]. Available: \url{https://arxiv.org/abs/2306.12338}
\BIBentrySTDinterwordspacing

\bibitem{10830749}
J.~Kim, T.-T.-H. Le, S.~Lee, and H.~Kim, ``Ethereum smart contracts vulnerabilities detection leveraging fine-tuning distilbert,'' in \emph{2024 International Conference on Platform Technology and Service (PlatCon)}, 2024, pp. 133--138.

\bibitem{ma2024combining}
W.~Ma, D.~Wu, Y.~Sun, T.~Wang, S.~Liu, J.~Zhang, Y.~Xue, and Y.~Liu, ``Combining fine-tuning and llm-based agents for intuitive smart contract auditing with justifications,'' \emph{arXiv preprint arXiv:2403.16073}, 2024.

\bibitem{1366126}
B.~Chess and G.~McGraw, ``Static analysis for security,'' \emph{IEEE Security \& Privacy}, vol.~2, no.~6, pp. 76--79, 2004.

\bibitem{10.1145/2976749.2978309}
\BIBentryALTinterwordspacing
L.~Luu, D.-H. Chu, H.~Olickel, P.~Saxena, and A.~Hobor, ``Making smart contracts smarter,'' in \emph{Proceedings of the 2016 ACM SIGSAC Conference on Computer and Communications Security}, ser. CCS '16.\hskip 1em plus 0.5em minus 0.4em\relax New York, NY, USA: Association for Computing Machinery, 2016, p. 254–269. [Online]. Available: \url{https://doi.org/10.1145/2976749.2978309}
\BIBentrySTDinterwordspacing

\bibitem{9072659}
J.~Chen, X.~Xia, D.~Lo, J.~Grundy, X.~Luo, and T.~Chen, ``Defining smart contract defects on ethereum,'' \emph{IEEE Transactions on Software Engineering}, vol.~48, no.~1, pp. 327--345, 2022.

\bibitem{10.1145/3194113.3194115}
\BIBentryALTinterwordspacing
S.~Tikhomirov, E.~Voskresenskaya, I.~Ivanitskiy, R.~Takhaviev, E.~Marchenko, and Y.~Alexandrov, ``Smartcheck: static analysis of ethereum smart contracts,'' in \emph{Proceedings of the 1st International Workshop on Emerging Trends in Software Engineering for Blockchain}, ser. WETSEB '18.\hskip 1em plus 0.5em minus 0.4em\relax New York, NY, USA: Association for Computing Machinery, 2018, p. 9–16. [Online]. Available: \url{https://doi.org/10.1145/3194113.3194115}
\BIBentrySTDinterwordspacing

\bibitem{9833721}
P.~Bose, D.~Das, Y.~Chen, Y.~Feng, C.~Kruegel, and G.~Vigna, ``Sailfish: Vetting smart contract state-inconsistency bugs in seconds,'' in \emph{2022 IEEE Symposium on Security and Privacy (SP)}, 2022, pp. 161--178.

\bibitem{10.5555/318773.318944}
T.~Ball, ``The concept of dynamic analysis,'' in \emph{Proceedings of the 7th European Software Engineering Conference Held Jointly with the 7th ACM SIGSOFT International Symposium on Foundations of Software Engineering}, ser. ESEC/FSE-7.\hskip 1em plus 0.5em minus 0.4em\relax Berlin, Heidelberg: Springer-Verlag, 1999, p. 216–234.

\bibitem{8952204}
M.~Mossberg, F.~Manzano, E.~Hennenfent, A.~Groce, G.~Grieco, J.~Feist, T.~Brunson, and A.~Dinaburg, ``Manticore: A user-friendly symbolic execution framework for binaries and smart contracts,'' in \emph{2019 34th IEEE/ACM International Conference on Automated Software Engineering (ASE)}, 2019, pp. 1186--1189.

\bibitem{10.1145/3395363.3404366}
\BIBentryALTinterwordspacing
G.~Grieco, W.~Song, A.~Cygan, J.~Feist, and A.~Groce, ``Echidna: effective, usable, and fast fuzzing for smart contracts,'' in \emph{Proceedings of the 29th ACM SIGSOFT International Symposium on Software Testing and Analysis}, ser. ISSTA 2020.\hskip 1em plus 0.5em minus 0.4em\relax New York, NY, USA: Association for Computing Machinery, 2020, p. 557–560. [Online]. Available: \url{https://doi.org/10.1145/3395363.3404366}
\BIBentrySTDinterwordspacing

\end{thebibliography}

\end{document}